\documentclass[11pt,a4paper]{article}
\usepackage{jheppub_kim}
\usepackage{url}
\usepackage{dcolumn,pifont}
\newcommand{\xmark}{\ding{55}}%  

\usepackage{pdflscape}
\usepackage{amsmath}
\usepackage{amssymb}
\usepackage{dcolumn}
\usepackage{bm}
\usepackage{color}
\usepackage{epsfig}
\usepackage{amsfonts}
\usepackage{graphicx}
\usepackage{subfigure}
\usepackage{dcolumn}
 
\newcommand{\sfrac}[2]{{\textstyle{#1\over#2}}}
\PassOptionsToPackage{linktocpage}{hyperref}
\def\case#1/#2{\textstyle\frac{#1}{#2}}

\newcommand{\be}{\begin{equation}}
\newcommand{\ee}{\end{equation}}
\newcommand{\ben}{\begin{eqnarray}}
\newcommand{\een}{\end{eqnarray}}

\def\sp{\sigma_+}
\def\udot{\dot{u}}
\def\ex{e_1{}^1}
\def\ey{e_2{}^2}

\def\y{\vartheta}
\def\z{\varphi}

\def\be{\begin{equation}}
\def\ee{\end{equation}}
\def\bea{\begin{eqnarray}}
\def\eea{\end{eqnarray}}

\begin{document}

\title{Dynamical analysis of
anisotropic scalar-field
cosmologies for a wide range of potentials}

\author[a]{Carlos R. Fadragas}

\author[b]{Genly Leon}

\author[c,b]{and Emmanuel N. Saridakis}

\affiliation[a]{Department of Physics, Universidad Central de
Las Villas, Santa Clara CP 54830, Cuba}

\affiliation[b]{Instituto de F\'{\i}sica, Pontificia Universidad  Cat\'olica
de Valpara\'{\i}so, Casilla 4950, Valpara\'{\i}so, Chile}
\affiliation[c]{Physics Division, National Technical University of Athens,
15780 Zografou Campus,  Athens, Greece}

\emailAdd{fadragas@uclv.edu.cu}

\emailAdd{genly.leon@ucv.cl}

\emailAdd{Emmanuel$_-$Saridakis@baylor.edu}

%\pacs{04.50.Kd, 98.80.-k, 95.36.+x}

\abstract{We perform a detailed dynamical analysis of anisotropic
scalar-field cosmologies, and in particular of the most significant
Kantowski-Sachs, Locally Rotationally Symmetric (LRS) Bianchi I and LRS
Bianchi III cases. We follow the new and
powerful method of  $f$-devisers, which allows us to perform the whole
analysis for a wide range of potentials. Thus, one can just substitute the
specific
potential form in the final results and obtain the corresponding behavior,
without the need of new calculations. We find a very rich behavior, and
amongst others the universe can result in isotropized solutions with
observables in agreement with observations, such as de Sitter,
quintessence-like, or stiff-dark energy solutions. In particular, all
expanding, accelerating, stable attractors  are  isotropic. 
Additionally, we
prove that as long as matter obeys the null energy condition, bounce behavior
is impossible. Finally, applying the general results to the well-studied
exponential  and
power-law potentials, we find that some of the general stable solutions
disappear. This feature may be an indication that such simple potentials
might restrict the dynamics in scalar-field
cosmology, opening the way to the introduction of more complicated
ones.}
%\pacs{98.80.-k, 95.36.+x, 04.50.Kd}

\maketitle

%\date{\today}
\newpage

\section{Introduction}

According to combined observations, the current observable universe is
homogeneous and isotropic at a great accuracy
\cite{Komatsu:2010fb,Barrow:1997sy,Bernui:2005pz}.
Although inflation is the most successful candidate for explaining such a
behavior, strictly speaking the problem is not fully solved, since in
the literature one usually starts from the beginning with a homogeneous
and isotropic Friedmann-Robertson-Walker (FRW) metric, examining the
evolution of fluctuations, instead of starting with an arbitrary
metric and
show that inflation is indeed realized and that the universe evolves towards
homogeneity and isotropy. However, this complete analysis is hard even for
numerical elaboration \cite{Gol:89}, and thus in order to extract  
analytical information one should impose one assumption more, namely to
consider anisotropic but homogeneous cosmologies. This class of geometries
\cite{Misner74} exhibits very interesting cosmological
features, both in inflationary and post-inflationary epochs
\cite{Peebles:1994xt}, and in these
lines  isotropization is a crucial issue, namely whether the universe can
result in isotropic solutions without the need of fine-tuning the model
parameters. Finally, the class of anisotropic geometries has recently gained
a lot of interest, under the light of the recently announced Planck Probe
results \cite{Ade:2013zuv}, in which some small anisotropic ``anomalies''
seem to
appear.

Amongst the various families of homogeneous but anisotropic geometries the 
most well-studied are the Bianchi type \cite{Ellis:1968vb} (see
\cite{Tsagas07} and references therein) and the Kantowski-Sachs
metrics \cite{chernov64,KS66,Burd:1988ss,Yearsley:1996yg}. Furthermore,
amongst the Bianchi subclass
the simplest but still very interesting geometries are the Bianchi I
\cite{Burd:1988ss,Yearsley:1996yg,Henneaux:1980ft,Muller:1985mr,
Berkin:1992ub,Aguirregabiria:1993pm,
Saha:2003xv,Barrow:2005dn,Clifton:2006kc,Leach2006,
Bandyopadhyay2007,Sharif:2009xa,
MartinBenito:2009qu,Aref'eva:2009vf} and  Bianchi III
\cite{Burd:1988ss,Louko:1987gg,Tikekar:1992ia,Christodoulakis:2006vi}.
Thus, in these geometries one can  analytically examine the rich behavior,
and incorporate additionally the matter content of the universe 
 \cite{Burd:1988ss,Yearsley:1996yg,Henneaux:1980ft,Muller:1985mr,
 Berkin:1992ub,Aguirregabiria:1993pm,Saha:2003xv,Clifton:2006kc,
 Barrow:2005dn,Leach2006,Bandyopadhyay2007,Sharif:2009xa,MartinBenito:2009qu,
Aref'eva:2009vf,Louko:1987gg,Tikekar:1992ia,Christodoulakis:2006vi,
Barrow:1994nt,Collins:1977fg,Weber:1984xh,Sahni1986,Demianski:1988js,
Campbell:1990uu,Chakraborty:1990jp,Mendes:1990eb,Nojiri:1999iv,Li:2003bq,
Chiou:2008eg,Leon:2010pu,Byland:1998gx,Ibanez:1995zs,Aguirregabiria:1996uh,
vandenHoogen:1996vc,Chimento:1998cf,vandenHoogen:1998cc,
Christodoulakis:1999ir,Christodoulakis:2000jm,Solomons:2001ef,
Middleton:2010bv,Czuchry:2012ad,Chimento:2012is,Keresztes:2013tua}.

On the other hand, there are now strong evidences that the observable
universe is accelerating \cite{Riess:1998cb,Perlmutter:1998np}, which led
scientists to follow
two directions in order to explain it. The first is to modify the
gravitational sector itself (see \cite{Nojiri:2006ri,Capozziello:2011et} for 
 reviews and references therein). The second is to introduce the concept of
dark energy
(see \cite{Copeland:2006wr} and references therein), which could be
the simple cosmological constant, a quintessence scalar field 
\cite{Ratra:1987rm,Wetterich:1987fm,Liddle:1998xm,Dutta:2009yb}, a phantom
field \cite{Caldwell:1999ew,Nojiri:2003vn,Onemli:2004mb,
Saridakis:2008fy}, or the
combination of
both these fields in a unified scenario named quintom
\cite{Feng:2004ad,Guo:2004fq,Zhao:2006mp,Setare:2008pz,
Cai:2009zp,Lazkoz:2006pa, Lazkoz:2007mx,Leon:2012vt}.  In all these
scalar-field-based scenarios a crucial quantity is the
field-potential, since it can radically affect the cosmological behavior.

In the present work we are interested in performing a dynamical analysis of
anisotropic scalar field cosmology. This phase-space and
stability examination
allows us to bypass the non-linearities and complications of the
cosmological equations, which prevent   complete analytical
treatments, obtaining a qualitative description of the global dynamics of
these scenarios, which is independent of the initial conditions and the
specific evolution of the universe. Moreover, in these asymptotic
solutions we calculate various observable quantities, such as the dark-energy
and total equation-of-state parameters, the deceleration parameter,
the various density parameters, and of course the isotropization measure.
However, in order to remain general, we extend beyond the usual procedure
\cite{Burd:1988ss,REZA,Copeland:1997et,Coley:2003mj,Leon2011,Gong:2006sp,
Setare:2008sf,Chen:2008ft,Gupta:2009kk,Matos:2009hf,
Copeland:2009be,Leyva:2009zz,
Farajollahi:2011ym,UrenaLopez:2011ur,Escobar:2011cz,
Escobar:2012cq,Xu:2012jf,Leon:2013qh}, and we
follow the method of the $f$-devisers introduced in \cite{Escobar:2013js},
which allows us to perform the whole analysis without the need of an  {\it{a
priori}}  specification of the potential. Thus, after such a general analysis
one can just substitute the specific potential form in the final results,
instead of having to repeat the whole dynamical elaboration from the start.
This general investigation appears for the first time in the literature and
it eliminates almost any calculation need from future relevant
investigations.

The manuscript is organized as follows: In section  \ref{Ancosm} we briefly
present the general features of scalar-field cosmology in anisotropic
geometry, introducing the kinematic variables and the various observables.
In section \ref{Dynamicalanalysis} we perform a complete dynamical analysis
for a wide range of potentials, and we discuss the isotropization and bounce
issues. In section \ref{exppotential} we apply the
obtained
results to the specific case of an exponential potential with a cosmological
constant, while in section \ref{powerlawpot} to the case of a power-law
potential. Finally, in section \ref{conclusions} we summarize the obtained
results.

\section{Anisotropic cosmology}
\label{Ancosm}

In this section we present the basic features of scalar-field cosmology in
anisotropic Locally Rotationally Symmetric (LRS) Bianchi I, LRS Bianchi III
and Kantowski-Sachs geometries.
The corresponding action takes the usual form, namely
\cite{Ratra:1987rm,Wetterich:1987fm,Liddle:1998xm,Dutta:2009yb}
\begin{equation}
   {\cal S}_{\text{metric}}=\int  {d} x^4
\sqrt{-g}\left[\frac{R}{2}+\frac{1}{2}g^{\mu\nu}
\partial_\mu\phi\partial_\nu\phi-V(\phi)
+ { \cal
L}^{(m)}\right],
\label{action}
 \end{equation}
where $g^{\mu\nu}$ is the anisotropic metric described in detail below (we
use the metric signature $(-1,1,1,1)$), $R$ is the Ricci scalar, Greek
indices run from $0$ to $3$, and we impose the standard convenient units in
which
$c=8\pi G=1$. In the above action we have added a canonical minimally coupled
scalar field
$\phi$, with $V(\phi)$ its potential, in which as usual one attributes the
dark-energy sector. Additionally, ${\cal L}^{(m)}$ accounts for the
matter content of the universe, which is assumed to  be an orthogonal perfect
fluid
with energy density $\rho_m$ and pressure $p_m$, and thus its
equation-of-state parameter is $w_m=p_m/\rho_m$, while its barotropic
index reads $\gamma\equiv w_m+1$. Finally, for simplicity
we do not consider an explicit cosmological constant, which if necessary
can arise from  a constant term in the potential $V(\phi)$. 

Let us make a comment here concerning the relation of anisotropic
geometry with the universe content. In order to have full consistency, the
consideration of anisotropic geometry would need the consideration of an
anisotropic matter, and a vector instead of a scalar field or the vector
coupled to the scalar field (similarly to models of anisotropic inflation
\cite{Watanabe:2009ct,Kanno:2009ei,Kanno:2010nr,Watanabe:2010fh,Emami:2010rm,
Dimopoulos:2010xq,Dimopoulos:2011ym,Do:2011zza} where a scalar inflaton
is coupled to a vector, or similarly to anisotropic Einstein-aether scenarios
\cite{Alhulaimi:2013sha}, that is a vector-tensor model
of gravitational Lorentz violation  \cite{Jacobson:2000xp,Jacobson:2008aj}).
However, since the
post-inflationary anisotropies are small, as a first approach one can neglect
the anisotropies of the various fluids at the post-inflationary universe,
and focus only on the geometrical anisotropies
\cite{Aguirregabiria:1993pm,Saha:2003xv,Clifton:2006kc,Leach2006,
Aref'eva:2009vf,Li:2003bq,
Leon:2010pu,Byland:1998gx,Ibanez:1995zs,Aguirregabiria:1996uh,
vandenHoogen:1996vc,Solomons:2001ef,Keresztes:2013tua,
Escobar:2012cq}. Therefore,  in this work we consider a scalar field, and
the matter to be homogeneous and isotropic.

Let us now present the cosmological application of the above general
scalar-field action in a background of anisotropic geometry. Without loss
of generality we focus on the most interesting and well-studied cases,
introducing an anisotropic metric of the form
\cite{Byland:1998gx}:  
\begin{equation}
 ds^2 = - N(t)^2 dt^2 + [\ex(t)]^{-2} dr^2
  + [\ey(t)]^{-2} [d\y^2 + S(\y)^2\,
   d\z^2],\label{metric}
\end{equation}
where $1/\ex(t)$ and $1/\ey(t)$ are the expansion scale factors.
%The frame vectors in coordinate form are written as
%\begin{eqnarray}
%  &&  \e_0 = N^{-1} \partial_t ,
%    \e_1 = \ex \partial_r
%        \nonumber\\
%    &&    \e_2 = \ey \partial_\y,  \
%        \e_3 = \ey/S(\y) \partial_\z.
%\end{eqnarray} \
This metric can describe three geometric classes, given by three
corresponding forms of $S(\y )$,
namely
\begin{eqnarray}
S(\y ) &=&\left\{
\begin{array}{l}
\sin {\y }\hspace{0.7cm}:\hspace{0.3cm}{\rm Kantowski-Sachs}, \\
\y \hspace{1.2cm}:\hspace{0.3cm}{\rm LRS \  Bianchi\  I}, \\
\sinh {\y }\hspace{0.5cm}:\hspace{0.3cm}{\rm LRS \ Bianchi \ III},
\end{array}
\right. \label{threegeom}
\end{eqnarray}
known respectively as Kantowski-Sachs, LRS Bianchi I and LRS Bianchi III
geometries.
{{ It proves convenient to express the above three cases in a unified
way  following \cite{Nilsson:1995ah}, which will later allow to perform
the dynamical analysis in a unified way too, using
a {\it{dominant}} normalization variable.
This approach was introduced in \cite{uggla1990}, where the LRS Bianchi
type IX  models were reduced to a regularized first-order system
of differential equations for some bounded gravitational variables.  
In particular,  the LRS Bianchi type I model naturally appears as a boundary
subset in LRS Bianchi type III, which appears as an invariant boundary of the
LRS Bianchi type VIII  models, 
%and in the Kantowski-Sachs case (for $Q=\pm
%1$, see definition
%\eqref{auxiliary}), 
which can be viewed as an invariant boundary of the LRS
Bianchi type IX models \cite{uggla1990,Nilsson:1995ah} (see
\cite{Goliath:1998mw,Carr:1999rv} for a similar global
dynamical analysis for spatially
self-similar, spherically symmetric,  perfect-fluid models, and
\cite{Coley:2000zw,Coley:2000yc}  
under the addition of a scalar field).

Following \cite{Nilsson:1995ah} we redefine the metric
\eqref{metric} as:
\begin{equation}\label{metricA}
 ds^2 = - N(t)^2 dt^2 + [\ex(t)]^{-2} dr^2
  + [\ey(t)]^{-2} [d\y^2 + k^{-1}\sin(\sqrt{k}\y)^2\,
   d\z^2].
\end{equation}
Henceforth, we can deal with the aforementioned three metrics as an
one-parameter family of metrics, where the choices $k=+1$ and $k=-1$
correspond respectively to Kantowski-Sachs and LRS Bianchi III, and
$k\rightarrow 0$ corresponds to Bianchi I. 
 }   }

We next consider relativistic fluid dynamics (see for instance
\cite{cargeellis73}) in such a geometry. For any given fluid 4-velocity
vector field $u^{\mu}$, the projection tensor 
$h_{\mu \nu}=g_{\mu \nu}+u_\mu u_\nu$ projects into the instantaneous
rest-space of a comoving observer 
(letters from the second half of the Greek alphabet denote covariant
spacetime indices). As usual we decompose the
  covariant derivative $\nabla_\mu u_\nu$ into its irreducible
parts
$$\nabla_\mu u_\nu=-u_\mu\dot u_\nu+\sigma_{\mu \nu}+\frac{1}{3}\Theta
h_{\mu \nu}-\omega_{\mu
\nu},$$ where $\udot_\mu$ is the acceleration vector defined as
$\udot_\mu=u^\nu\nabla_{\nu}u_{\mu}$ (the dot denotes a derivative
with respect to $t$ for the choice of the 4-velocity vector  field ${\bf
u}=\partial_t$),  $\sigma_{\mu \nu}$ is the symmetric and
trace-free
shear tensor ($\sigma_{\mu \nu} = \sigma_{(\mu \nu)}$,
$\sigma_{\mu \nu}\,u^\nu = 0$, $\sigma^\mu{}_{\mu}= 0$), and
$\omega_{\mu \nu}$ is the antisymmetric vorticity tensor
($\omega_{\mu \nu} = \omega_{[\mu \nu]}$, $\omega_{\mu \nu}\,u^\nu
= 0$). 
Moreover, in the above expression we have also introduced
 the volume expansion scalar
$\Theta= \nabla_\mu u^\mu$,
 which as usual defines a representative
length scale $\ell$
along the flow lines \cite{EU}:
\begin{equation}
\Theta \equiv\sfrac{3 u^\mu \nabla_\mu \ell}{\ell}=\sfrac{3 \dot\ell}{\ell},
 \label{thetaelldef}
\end{equation}
 describing the volume expansion or contraction.  That is, $\ell$ is a form
of average ``scale factor'', while $\Theta$ is a form of average Hubble
parameter. In particular, it is standard to define the Hubble parameter as
$H=\Theta/3$ \cite{EU}, and thus it becomes clear that in FRW geometries
$\ell$ coincides with the usual scale factor. Furthermore, one can show
that these kinematic fields are related through
\cite{cargeellis73,EU}
\begin{align}
&\sigma_{\mu \nu}:=\dot u_{(\mu} u_{\nu)}+\nabla_{(\mu}
u_{\nu)}-\frac{1}{3} \Theta h_{\mu \nu}\\
& \omega_{\mu \nu}:= -u_{[\mu}\dot u_{\nu]}-\nabla_{[\mu}
u_{\nu]}.
\end{align}

It proves more convenient to work into the synchronous
temporal gauge, where we can set $N$ to
any positive function of $t$, with the simplest choice being $N=1$.
Therefore, one obtains the following constraints on kinematical
variables:
\begin{equation}
   \sigma^\mu_\nu = \text{diag}(0,-2\sp,\sp,\sp),\ \
    \omega_{\mu \nu} =0,
    \end{equation}
     where we have defined
     \begin{equation}
     \label{sigmaplus}
      \sp\equiv\frac13\frac{  d}{ {d} t}\left[\ln
      \frac{\ex}{\ey}\right].
       \end{equation}
Thus, the Hubble parameter can be written in
terms of
$\ex$ and $\ey$ as
    \begin{equation}
\label{Hubblepar}
    H=-\frac13\frac{  d}{ {d} t}\left[\ln \ex
(\ey)^2\right],
\end{equation}
while one can additionally define the Gauss curvature of the 3-spheres
 \cite{EU} as
 \begin{equation}
\label{Gausscurv}
    K = (\ey)^2.
\end{equation}

Finally, in order to relate the above analysis to observations, one defines 
  various density parameters, namely \cite{REZA}: the
curvature one
\begin{equation}
\label{Omegas1}   \Omega_k\equiv -\frac{k K}{3H^2},
\end{equation}
the matter one
\begin{equation}
\label{Omegas2}
   \Omega_{m}\equiv\frac{\rho_m}{3 H^2},
\end{equation}
the shear  one
\begin{equation}
\label{Omegas3}
\Omega_\sigma\equiv\left(\frac{\sp}{H}\right)^2,
\end{equation}
and the dark-energy one, which is just the  scalar-field one,
\begin{equation}
\Omega_{DE}\equiv\Omega_{\phi}\equiv\frac{\frac{1}{2}\dot\phi^2+V(\phi)}{
3H^2 } ,
\label{Omegas4}
\end{equation}
where we have assumed that   both matter and   scalar field are
homogeneous, that is they depend only on time.
 %satisfying $\Omega_k+\Omega_m+\Omega_{\phi}+\Omega_\sigma=1$.
Furthermore,   we can
straightforwardly write the deceleration parameter as
\begin{equation}
q=-1-\frac{\dot{H}}{H^2},
\label{qKS}
\end{equation}  the dark energy equation-of-state
parameter $w_{DE}$ as
\begin{equation}
w_{DE}\equiv w_\phi= \frac{\dot \phi^2-2 V(\phi)}{\dot \phi^2+2 V(\phi)}
\label{wDEKS},
\end{equation}
while it proves convenient to define as usual the ``total'' equation-of-state
parameter   
\begin{equation}
w_{tot}\equiv-1-\frac{2\dot{H}}{3H^2}=\frac{2q-1}{3}.
\label{wDtot}
\end{equation}

We mention here that in anisotropic scalar-field cosmology there is a
discussion on the precise definition of the dark-energy sector, namely
whether it should include the shear term along the scalar-field one. Such a
question is related to the observational capabilities, and in particular
on whether one can measure the shear and the scalar-field energy densities
separately. Since $\Omega_\sigma$ can indeed be estimated separately by
measuring the luminosity distance in different directions (if
statistically there is a significant anisotropy), one can define the average
deviation from $\Omega_{m}$ as $\Omega_{DE}\equiv\Omega_{\phi}$, while the
anisotropic component will be just $\Omega_\sigma$ 
\cite{Grande:2011hm}. Therefore, in this work we define the dark-energy
sector to be the scalar-field alone as usual, and in the various Tables we
provide $\Omega_{DE}$,  $\Omega_\sigma$  and $w_{DE}$  separately.

In summary, we have now all the required information to proceed to a
detailed investigation of scalar-field anisotropic cosmology, namely the
action (\ref{action}), the kinematic variables (\ref{sigmaplus}),
(\ref{Hubblepar}) and (\ref{Gausscurv}), and the observables
(\ref{Omegas1})-(\ref{wDtot}). In the following subsection
we extract the cosmological equations for the three geometries
of (\ref{threegeom}), namely the Kantowski-Sachs, the LRS Bianchi I and
the LRS Bianchi III ones in a unified way \cite{Nilsson:1995ah}.

\subsection{Unified approach for the field equations}

{{
Taking the variation of the action (\ref{action}) for the 1-parameter family  of metrics \eqref{metricA} leads to 
\begin{eqnarray}
&&
3H^2+k K=3\sp^2+{\rho_m}+\frac{1}{2}\dot\phi^2+V(\phi)
\label{KSsfpfGauss}\\
 &&-3(\sp+H)^{2}-2\dot\sp-2\dot H
-k K=(\gamma-1)\rho_m
+\frac{1}{2}\dot\phi^2-V(\phi)\label{KSsfpfvaryingaction2}\\
&&-3\sp^{2}+3\sp H -3H ^{ 2}+\dot{\sigma}_+-2\dot H=
(\gamma-1)\rho_m+\frac{1}{2}\dot\phi^2-V(\phi),
\label{KSsfpfvaryingaction3}
\end{eqnarray}
where we have used the Ricci-scalar relation
$  R= 12H ^{2}+6\sp^{2}+6\dot H +2 k K$. 

In the above equations the choice $k=+1$ corresponds to Kantowski-Sachs metric \cite{Burd:1988ss,Collins:1977fg,Weber:1984xh,Sahni1986,Demianski:1988js,
Campbell:1990uu,
Chakraborty:1990jp,Mendes:1990eb,Nojiri:1999iv,Li:2003bq,Chiou:2008eg,
Leon:2010pu}, $k=-1$ corresponds to LRS Bianchi III \cite{Burd:1988ss,Louko:1987gg,Tikekar:1992ia,Christodoulakis:2006vi} and $k=0$ corresponds to LRS Bianchi I \cite{Burd:1988ss,Henneaux:1980ft,Muller:1985mr,Berkin:1992ub,
Aguirregabiria:1993pm,
Saha:2003xv,Barrow:2005dn,Clifton:2006kc,Leach2006,
Bandyopadhyay2007,Sharif:2009xa,
MartinBenito:2009qu,Aref'eva:2009vf}.

Additionally, the evolution of the Gauss curvature is given by
 \begin{equation}
 \dot {K}=-2 (\sp +H) K\label{KSsfpfpropK},
\end{equation}
while the evolution equation for  $\ex$ writes as   \cite{Coley:2008qd} 
 \begin{equation} \dot{e}_1{}^1 =-
\left(H-2\sp\right)\ex\label{KSsfpfaux}.
\end{equation}
Combining \eqref{KSsfpfvaryingaction2} and
\eqref{KSsfpfvaryingaction3} we obtain the shear evolution equation
\begin{equation}
\label{KSsfpfevolanisotropies}
\dot{\sigma}_+=-3 H \sp-\frac{k K}{3}.
\end{equation}
Furthermore, relations
 \eqref{KSsfpfGauss}, \eqref{KSsfpfvaryingaction2},
\eqref{KSsfpfvaryingaction3} and \eqref{KSsfpfevolanisotropies} give
the Raychaudhuri equation
\begin{equation}
\label{KSsfpfRaych1}
\dot H
=-H^2-2\sp^2-\frac{1}{6}\left(3 \gamma -2\right)\rho_m-\frac{1}{3}
\dot\phi^2+\frac{1}{3}V(\phi).
\end{equation} where we have used the matter equation of state $p_m=(\gamma-1)\rho_m.$

Finally, variation of the action (\ref{action}) with respect to the matter
energy density and the scalar field, gives respectively the corresponding
evolution equations, namely
\begin{equation}\label{KSsfpfevolmatter}
\dot\rho_m =-3 H \gamma\rho_m
\end{equation}
and
\begin{equation}\label{KSsfpfKG}
\ddot\phi=-3 H\dot\phi-\frac{dV(\phi)}{d\phi},
\end{equation}
where the last one can equivalently be written as
$\dot{\rho}_\phi+3H(1+w_\phi)\rho_{\phi}=0$, since
$\rho_{\phi}=\dot{\phi}^2/2+V(\phi)$ and $w_\phi$ is given by (\ref{wDEKS}).

In summary, the relevant cosmological equations for scalar-field cosmology
for the 1-parameter family of metrics \eqref{metricA} are:  the Gauss
constraint or Hamiltonian constraint in the Hamiltonian formulation (which reduces to the Friedmann equation in the isotropic case)
\eqref{KSsfpfGauss}, the evolution equation for 2-curvature $K$
\eqref{KSsfpfpropK}, the evolution equation for the anisotropies
\eqref{KSsfpfevolanisotropies}, the Raychaudhuri equation
\eqref{KSsfpfRaych1}, and the matter and scalar-field
evolution equations \eqref{KSsfpfevolmatter} and \eqref{KSsfpfKG},  (note
that the evolution equation for $\ex$ \eqref{KSsfpfaux} decouples from the
rest, allowing for a dimensional reduction of the dynamical system). Lastly,
the various density parameters, the   deceleration
parameter and the dark-energy and total equation-of-state parameters are  
given by
(\ref{Omegas1})-(\ref{wDtot}).
}}
%{\bf{We close this section making the comment that this approach is suitable
%for 
%investigating more general classes
%like the ones presented in table 1 of \cite{Nilsson:1995ah} which contains LRS
%Bianchi I, LRS Bianchi III and Kantowski-Sachs perfect fluid models as
%special cases by introducing the parametrization for the line element 
%\begin{align}\label{metricB}
%& d\tilde{s}^2 = e^{-2 f r}ds^2\nonumber \\ 
%&= e^{-2 f r}\left(-  dt^2 + [\ex(t)]^{-2} dr^2
%  + [\ey(t) e^{a r}]^{-2} [d\y^2 + k^{-1}\sin(\sqrt{k}\y)^2\,
%   d\z^2]\right).
%\end{align} where 
%$f, a, k$
%are parameters describing the symmetry groups of the various models ($a k=0$).
%Additionally, for analyzing the possible recollapsing models, like are the
%Kantowski-Sachs models, it is useful to use a {\it{dominant}} normalization
%variable. This idea was introduced in \cite{uggla1990}, where the LRS
%Bianchi type IX perfect fluid models were reduced to a regularized
%first-order
%system of differential equations for some bounded gravitational variables.
%}}

\section{Local dynamical analysis for a wide range of potentials}
\label{Dynamicalanalysis}

In the previous section we presented scalar-field cosmology in three
anisotropic geometries, namely the Kantowski-Sachs, the LRS Bianchi I, and
the
LRS Bianchi III ones using a unified approach. In this section we perform the
full phase-space
analysis of these scenarios, however in order to remain general we do not
specify the scalar potential form, keeping it arbitrary.

As it is known, in order to perform the dynamical analysis
of a cosmological model we have to introduce suitable auxiliary
variables in order to transform the cosmological equations into an autonomous
dynamical system
\cite{Burd:1988ss,REZA,Copeland:1997et,Coley:2003mj,Leon2011}, that
is in a form  $\textbf{X}'=\textbf{f(X)}$, where $\textbf{X}$ is the
column vector of the auxiliary variables, $\textbf{f(X)}$
is the column vector of the autonomous equations, and primes
denote derivatives with respect to $\ln a$. The
critical points $\bf{X_c}$ are extracted by the requirement $\bf{X}'=0$,
and in order to determine their stability properties  we expand around
$\bf{X_c}$ as $\bf{X}=\bf{X_c}+\bf{U}$, with $\textbf{U}$ the column vector
constituted by the perturbations of the auxiliary variables. Therefore, up
to first order we obtain $\textbf{U}'={\bf{\Xi}}\cdot \textbf{U}$, where  the
coefficients of the perturbation equations are contained inside the matrix
${\bf {\Xi}}$, and then the type and stability of each critical point is
determined by the eigenvalues of ${\bf {\Xi}}$ (stable (unstable) point
for eigenvalues with negative (positive) real parts, or saddle point
for eigenvalues with real parts of different sign).

In order to follow the above procedure and in order to handle the involved
differentiations, it is necessary to determine a specific potential form
$V(\phi)$ of the scalar field $\phi$, the parameters of which will
determine the aforementioned eigenvalues
\cite{Burd:1988ss,REZA,Copeland:1997et,Coley:2003mj,Leon2011,Gong:2006sp,
Setare:2008sf,
Chen:2008ft,Gupta:2009kk,Matos:2009hf,
Copeland:2009be,Leyva:2009zz,
Farajollahi:2011ym,UrenaLopez:2011ur,Escobar:2011cz,Escobar:2012cq,
Xu:2012jf,Leon:2013qh}. The disadvantage  of this procedure
is that for each different potential one must repeat all the calculations
from the beginning. Therefore, it would be very helpful to develop an
extended method that could handle the potential differentiations in
a unified way, without the need of any {\it{a priori}} specification. This is
exactly the method of $f$-devisers introduced in \cite{Escobar:2013js}  and
applied in \cite{delCampo:2013vka} for scalar-field FRW cosmologies in the
presence of a Generalized Chaplygin Gas.
 \begin{table*}[ht]
\begin{center}
\resizebox{\columnwidth}{!}{
\begin{tabular}{@{\hspace{4pt}}c@{\hspace{14pt}}c@{\hspace{
14pt}}c}
\hline
\hline\\[-0.3cm]
Potential & References& $f(s)$  \\[0.1cm] 
\hline\\[-0.2cm]
%%%%%%%%
$V(\phi)=V_{0}e^{-\lambda\phi}+\Lambda$& 
\cite{Yearsley:1996yg,Pavluchenko:2003ge,Cardenas2003} &
  $-s(s-\lambda)$\\[0.2cm]
$V(\phi)=V_{0}\left[e^{\alpha\phi}+e^{\beta\phi}\right]$&
\cite{Barreiro:1999zs,Gonzalez:2007hw,Gonzalez:2006cj} &
$-(s+\alpha)(s+\beta)$ \\[0.2cm]
$V(\phi)=V_{0}\left[\cosh\left( \xi \phi \right)-1\right]$
&\cite{Ratra:1987rm,Wetterich:1987fm,Matos:2009hf,Copeland:2009be,
Leyva:2009zz, Pavluchenko:2003ge,delCampo:2013vka,
Sahni:1999qe,Sahni:1999gb, Lidsey:2001nj,
Matos:2000ng} &  $-\frac{1}{2}(s^2-\xi^2)$ \\[0.2cm]
$V(\phi)=V_{0}\sinh^{-\alpha}(\beta\phi)$&
\cite{Ratra:1987rm,Wetterich:1987fm,Copeland:2009be,Leyva:2009zz,
Pavluchenko:2003ge,Sahni:1999gb, UrenaLopez:2000aj}  &
$\frac{s^2}{\alpha}-\alpha\beta^2$ \\[0.4cm]
\hline \hline
\end{tabular}}
\end{center}
\caption{\label{fsform} The function  $f(s)$ for the most common
quintessence potentials \cite{Escobar:2013js}.}
\end{table*}

In this extended dynamical analysis method, one introduces two new   
dynamical variables $s$ and $f$ as
\begin{eqnarray}
\label{sdef}
s&\equiv&-\frac{V_\phi(\phi )}{V(\phi
)},\\
f&\equiv&\frac{V_{\phi\phi}(\phi
   )}{V(\phi )}-\frac{V_\phi(\phi )^2}{V(\phi )^2},
\label{fdef}
\end{eqnarray}
where the subscript ``$\phi$'' denotes differentiation with respect to
$\phi$.
In principle $f$ can be expressed as an explicit function of $s$, that is
$f=f(s)$. Therefore, following the above procedure, we can transform our
cosmological system into a closed dynamical system for a set of normalized,
auxiliary, variables and $s$. Such a procedure is possible for a wide range
of potentials, especially for the usual ansatzes of the cosmological
literature, where it
results to very simple forms for $f(s)$, as can be seen in Table \ref{fsform}
(note that for the single exponential potential the $s$-variable is not
required, since it is a constant and thus $f$ is automatically zero).
However, we
mention that the method cannot be applied to arbitrary potentials, but only
to those that allow for a solution of $f$ in terms of
$s$. For instance, in some specific forms such is the logarithmic
$V(\phi)\propto \phi ^p \ln ^q(\phi )$ \cite{Barrow:1995xb} and the
generalized exponential one $V(\phi)\propto \phi^n e^{-\lambda\phi^m}$
\cite{Parsons:1995ew}, used
in the inflationary context, $f$ cannot be expressed as a
single-valued
function of $s$. In these cases one should apply asymptotic 
techniques in order to extract the dominant branch at large $\phi$-values as
in \cite{Barrow:1995xb,Parsons:1995ew}. However, in general, for a wide
range of potentials the introduction of
the
variables $f$ and $s$ adds an extra direction in the phase-space, whose
neighboring points correspond to ``neighboring'' potentials. Therefore, after
the
general analysis has been completed, the substitution of the specific $f(s)$
for the desired potential gives immediately the specific results, through a
form of intersection of the extended phase-space.

On the other hand, when $f(s)$ is given we can straightforwardly reconstruct
the corresponding potential-form starting with
\begin{eqnarray}
&& \frac{ds}{d\phi}=-f(s)
\label{ds-dphi}\\
&& \frac{dV}{d\phi}=-s V,
\label{dV-dphi}
\end{eqnarray}
which lead to
\begin{eqnarray}
\phi(s)&=&\phi_0-\int_{s_0}^s \frac{1}{f(K)} \, dK\label{quadphi}\\
V(s)&=&e^{\int_{s_0}^s \frac{K}{f(K)} \, dK} \bar{V}_0\label{quadV},
\end{eqnarray}
 with the integration constants satisfying 
$\phi(s_0)=\phi_0$, $V(s_0)=\bar{V}_0$. Note that relations
(\ref{quadphi}) and (\ref{quadV}) are always valid, giving the
potential in an implicit form. However, for the usual cosmological cases of
Table \ref{fsform}  we can additionally eliminate  $s$  between
\eqref{quadphi} and \eqref{quadV}, and write the potential explicitly as
$V=V(\phi)$.

As a specific example let us take the inverse hyperbolic-sine
potential. In this case, substitution of the corresponding $f(s)$ of Table
\ref{fsform} in \eqref{quadphi} and \eqref{quadV} gives
$\phi(s)=[\beta  \phi_0+\coth ^{-1}(\alpha  \beta
   /s)-\coth ^{-1}( \alpha  \beta
   /s_0)]/\beta$
 and $V(s)=\bar{V}_0 \left(s^2-\alpha ^2 \beta
^2\right)^{\alpha /2}
   \left(s_0^2-\alpha ^2 \beta ^2\right)^{-\alpha /2}$. Thus, using the
parameter values $\bar{V}_0=V_0 \sinh ^{-\alpha }(\beta
\phi_0)$ and $s_0=-\alpha  \beta  \coth (\beta \phi_0)$ we obtain
$V(\phi)=V_0 \sinh^{-\alpha }[\beta(\phi -2 \phi_0)]$ and the
potential $V=V_{0}\sinh^{-\alpha}(\beta\phi)$ is recovered at
$\phi_0= 0$.

Finally, note that the $f$-devisers method  
allows also to reconstruct a scalar field potential from a model with
stable fixed points. In particular, choosing a function $f(s)$ with
the requested properties (existence of minimum, intervals of monotony,
differentiability) to have late-time stable attractors,
one uses \eqref{quadphi} and \eqref{quadV} to explicitly obtain  
$V(\phi)$. This is similar to the superpotential construction
method  \cite{Aref'eva:2009xr}, which allows for the construction of stable
kink-type solutions in scalar-field cosmological models, starting from the
dynamics, and specifically for the Lyapunov stability.

\subsection{Kantowski-Sachs scalar-field cosmology}\label{KSphase_space}

{ For Kantowski-Sachs scalar-field cosmology the cosmological equations
are
\eqref{KSsfpfGauss},\eqref{KSsfpfpropK},\eqref{KSsfpfevolanisotropies},
\eqref{KSsfpfRaych1},\eqref{KSsfpfevolmatter},\eqref{KSsfpfKG} for the choice
$k=+1$. In order to analyze possible recollapsing models, such as the
Kantowski-Sachs ones, it is useful to use a {\it{dominant}} normalization
variable. This idea was introduced in \cite{uggla1990}, where the LRS
Bianchi type IX perfect-fluid models were reduced to a regularized
first-order
system of differential equations for some bounded gravitational variables.
Thus,  in order to transform this
cosmological system  in its
autonomous form we introduce the auxiliary variables:
\begin{equation}
Q=\frac{H}{D},\ \ \ 
\Sigma=\frac{\sp}{D}, \ \ \  x=\frac{\dot \phi}{\sqrt{6} D},\ \ \ 
y=\frac{\sqrt{V}}{\sqrt{3}D},\ \ \ 
 z=\frac{\rho_m}{3 D^2}, \ \ \ {\cal K}=\frac{k K}{3 D^2}, \label{auxiliary}
\end{equation}
where $D=\sqrt{H^2+\frac{1}{3} K}$ is the  {\it{dominant}} normalization
variable. }
Furthermore, we introduce the variable $\tau$ through
$ {d}\tau=D {d} t$,
and from now on primes will denote derivatives with respect to
$\tau$ (in the case of flat geometry $\tau$ becomes just $\ln a$). We stress
here that apart from the above variables, we will use
the
two variables $s$ and $f$ defined in (\ref{sdef}), (\ref{fdef}), in order
to handle the a wide range of potential $V(\phi)$.

In terms of these auxiliary variables the first Friedmann equation
\eqref{KSsfpfGauss} becomes
$ x^2+y^2+z+\Sigma^2=1$,  the $D$-definition becomes
$ Q^2+{\cal K}=1$, while from  (\ref{auxiliary}) we see that
$y\geq 0,\, z\geq 0$ and ${\cal K}\geq 0$. Therefore, we
conclude that the auxiliary variables obey $-1\leq Q\leq1$, $-1\leq
\Sigma\leq1$, $-1\leq x\leq1$, $0\leq y\leq1$, $0\leq z\leq1$, $0\leq
{\cal K}\leq1$, while $s$ is un-constrained. { By  construction the
normalized variables are bounded in the above intervals, but $s$ can take
values over the real line. Thus, a complete investigation should
include an analysis of the behavior at infinity,
using the Poincar\'e
projection method,  as in \cite{Byland:1998gx,Xu:2012jf}, however for
simplicity we do not proceed to
such a detailed analysis in the present work.}

Now, concerning the physical meaning of the auxiliary variables, we
deduce that the sign of $Q$ (which is the sign of $H$) determines whether the
universe is
expanding or contracting, ${\cal K}$  which is related with $\Omega_k$
defined in \eqref{Omegas1} by $\Omega_k=-\frac{{\cal K}}{Q^2}$)  determines
the curvature (${\cal K}=0$
corresponds to flat universe), while $\Sigma$ determines the anisotropy level
($\Sigma= 0$ corresponds to isotropization).

Additionally, in terms of the
auxiliary
variables the various density parameters, the deceleration
parameter and the dark-energy and total
equation-of-state
parameters, defined in
(\ref{Omegas1})-(\ref{wDtot}), read:
\begin{align}
 & \Omega_k=1-\frac{1}{Q^2},\nonumber\\
 &  \Omega_{m}=\frac{1-\left(x^2+y^2+\Sigma^2\right)}{Q^2},\nonumber\\
 &\Omega_{DE}\equiv\Omega_{\phi}=\frac{x^2+y^2}{Q^2},\nonumber\\
&\Omega_\sigma=\left(\frac{\Sigma}{Q}\right)^2,\nonumber\\
&q=-\frac{3 x^2 (\gamma -2)+3
   \gamma  \left(y^2+\Sigma ^2-1\right)-6 \Sigma ^2+2}{2 Q^2}, \nonumber\\
 &w_{DE}=-1+\frac{2x^2}{x^2+y^2},\nonumber\\
 &w_{tot}=\frac{x^2 (\gamma -2)+\gamma
   \left(y^2+\Sigma ^2-1\right)-\Sigma ^2+1}{\Sigma ^2-1}.
\label{OmegasKSaux}
\end{align}

In summary, using the dimensionless auxiliary variables
(\ref{auxiliary}), along with the two constraints, we result to the
five-dimensional dynamical system:
\begin{align}
& Q'= -\frac{1}{2} \left(Q^2-1\right) \left[3 (\gamma -2) x^2+2 (Q-3
\Sigma ) \Sigma 
+3 \gamma  \left(y^2+\Sigma
^2-1\right)+2\right],\label{KSQ}\\
& \Sigma'= -\frac{3}{2} Q (\gamma -2) \Sigma
^3+\left(1-Q^2\right) (\Sigma ^2-1) -\frac{3}{2} Q
\left[(\gamma -2) x^2+\left(y^2-1\right) \gamma
   +2\right] \Sigma ,\label{KSSigma}\\
& x'= -\frac{3}{2} Q (\gamma -2) x^3+\frac{1}{2} x \left\{2 \Sigma
-Q \left[2 (Q-3 \Sigma ) \Sigma +3 \gamma \left(y^2+\Sigma
   ^2-1\right)+6\right]\right\} +\sqrt{\frac{3}{2}} s y^2,\label{KSx}\\
& y'=\frac{1}{2} y \left\{-\sqrt{6} s x+2 \Sigma +Q \left[-3
(\gamma -2) x^2 -3 (\gamma -2) \Sigma ^2+3 \gamma -2 Q \Sigma
   \right]\right\}-\frac{3}{2} Q y^3 \gamma,\label{KSy}\\
& s'=-\sqrt{6} x f(s), \label{KSs}
\end{align}
defined on the phase space 
\begin{align}
& \{(Q,\Sigma,x,y,s)\in\mathbb{R}^5:-1\leq Q\leq1,-1\leq
\Sigma\leq1,-1\leq x\leq1, 0\leq y\leq1, \nonumber \\ & \ \ \ \ \ \ \ \ \ \ \
\ \  0\leq  x^2+y^2+\Sigma^2\leq 1\}.
\end{align}

\begin{table*}[ht]
\begin{center}
\resizebox{\columnwidth}{!}
{
\begin{tabular}{|c|c|c|c|c|c|c|c|}
\hline \hline
  Name & $Q$ &$\Sigma$ & $x$ & $y$  & $s$ & Existence & Stability \\
\hline\hline
$P_1^\epsilon$ & $\epsilon$ & $\epsilon$ & $0$ & $0$ & $s_c$ &
always &  unstable (stable)   \\[0.2cm] \hline
$P_2^\epsilon$ & $\epsilon$ & $-\epsilon$ & $0$ & $0$ & $s_c$ &
always &     unstable (stable) \\[0.2cm] \hline
 $P_3^\epsilon$ & $\frac{2\epsilon}{4-3\gamma}$ &
$\frac{(3\gamma-2)\epsilon}{4-3\gamma}$ &
$0$ & $0$ & $s_c$ & $0\leq \gamma\leq \frac{2}{3}$ &   saddle \\ \hline
$P_4^\epsilon$ & $\epsilon$ & $0$ & $0$ & $0$ & $s_c$ &
always &   saddle \\[0.2cm] \hline
$P_5^\epsilon$ & $\epsilon$ & $0$ & $0$ & $1$ & $0$ & always &   stable
(unstable) \\
 &    &   &   &   &  &    &  for $f(0)>0.$
\\[0.2cm] \hline
$P_6^\epsilon$ & $\frac{\epsilon}{2}$ & $-\frac{\epsilon}{2}$ & $0$ &
$\frac{\sqrt{3}}{2}$ & $0$ & always  &   saddle \\[0.2cm] \hline
$P_7^+$ & $1$ & $0$ & $-1$ & $0$ & $0$ & $s^*=0$ &  saddle
 for $f'(0)<0$ 
\\  
 &  &   &   &   &   &   &   
 unstable  for   $f'(0)>0$ 
\\[0.2cm] \hline
$P_7^-$ & $-1$ & $0$ & $-1$ & $0$ & $0$ & $s^*=0$  & stable 
for $f'(0)<0$ 
\\
  &   &  &   &   &   &   &saddle 
for $f'(0)>0$ 
\\[0.2cm] \hline
$P_8^+$ & $1$ & $0$ & $1$ & $0$ & $0$ & $s^*=0$  &
unstable  
for $f'(0)<0$  
\\ 
  &  &   &   &   &   &   &
 saddle 
for  $f'(0)>0$ 
\\[0.2cm] \hline
$P_8^-$ & $-1$ & $0$ & $1$ & $0$ & $0$ &$s^*=0$  &
 stable for  $f'(0)>0$ 
\\ 
  &   &   &   &   &   &   &
saddle for  $f'(0)<0$  
\\[0.2cm] \hline
$P_9^\epsilon(s^*)$ & $\epsilon\frac{2}{4-3 \gamma }$ & $-\epsilon\frac{2-3
\gamma }{4-3 \gamma }$  &
$\epsilon\frac{\sqrt{6} \gamma }{(4-3 \gamma ) s^*}$ &
   $\sqrt{\frac{6(2-\gamma ) \gamma }{(4-3 \gamma )^2 \left({s^*}\right)^2}}$
& $s^*$ &
$0<\gamma\leq \frac{2}{3},$ &
\\
 &    &   &   &   &  &  $\left({s^*}\right)^2\geq
\frac{\gamma}{1-\gamma}$ &  saddle  \\[0.2cm] \hline
$P_{10}^\epsilon(s^*)$ & $\frac{2\epsilon
\left[\left({s^*}\right)^2+1\right]}{\left({s^*}\right)^2+4}$ &
$\frac{\epsilon\left[\left({s^*}\right)^2-2\right]}{\left({s^*}\right)^2+4}$
&
$\frac{\sqrt{6}\epsilon
   s^*}{\left({s^*}\right)^2+4}$ & $\frac{\sqrt{6}
\sqrt{\left({s^*}\right)^2+2}}{\left({s^*}\right)^2+4}$ & $s^*$
& $0<\left({s^*}\right)^2\leq 2$ &   saddle \\[0.2cm] \hline
$P_{11}^\epsilon(s^*)$ & $\epsilon$ & $0$ & $\frac{\epsilon s^*}{\sqrt{6}}$ &
$\sqrt{1-\frac{\left({s^*}\right)^2}{6}}$ & $s^*$ & $0<\left({s^*}\right)^2\leq
6$ &   stable
(unstable) for\\
 &    &   &   &   &  &    &  $0<\gamma\leq \frac{2}{3},
-\sqrt{3\gamma}<s^*<0, f'(s^*)<0$  or
\\
 &    &   &   &   &  &    &   $0<\gamma\leq \frac{2}{3},
0<s^*<\sqrt{3\gamma}, f'(s^*)>0$ or
\\[0.2cm]
 &    &   &   &   &  &    &   $\frac{2}{3}<\gamma\leq 2, -\sqrt{2}<s^*<0,
f'(s^*)<0$ or
\\
 &    &   &   &   &  &    &    $\frac{2}{3}<\gamma\leq 2, 0<s^*<\sqrt{2},
f'(s^*)>0.$
\\
 &    &   &   &   &  &    &   saddle otherwise
\\[0.2cm] \hline
$P_{12}^\epsilon(s^*)$ &  $\epsilon$ & $0$ & $\frac{\sqrt{3}
\gamma\epsilon}{\sqrt{2}s^*}$ & $\sqrt{\frac{3}{2}} \sqrt{\frac{(2-\gamma )
\gamma
}{\left({s^*}\right)^2}}$  & $s^*$ & $\left({s^*}\right)^2\geq 3\gamma,$  &  stable
(unstable) for \\
 &    &   &   &   &  &  $s^*\neq 0$  &  $0<\gamma<\frac{2}{3},\,
s^*<-\sqrt{3\gamma},f'(s^*)<0,$ or 
\\
 &    &   &   &   &  &    &   or $0<\gamma<\frac{2}{3},\,
s^*>\sqrt{3\gamma},f'(s^*)>0.$
\\
 &    &   &   &   &  &    &   saddle otherwise
\\[0.2cm] \hline
$C_-(s^*)$ & $-1$ & $\cos u$ & $\sin u$ & $0$ & $s^*$ & always &
stable for \\
&    &   &   &   &  &    & $0\leq \gamma<2, 0<u<\pi,$
\\[0.2cm]  
  &   &   &   &   &   &   & $s^*>-\sqrt{6} \csc (u),
f'\left(s^*\right)>0$ or \\
 &    &   &   &   &  &    & $0\leq \gamma<2,\pi<u< 2\pi,$
\\  
  &   &   &   &   &   &   & $s^*<-\sqrt{6} \csc
(u),
f'\left(s^*\right)<0$  \\
 &    &   &   &   &  &    &   saddle otherwise
 \\[0.2cm] \hline
$C_+(s^*)$ & $1$ & $\cos u$ & $\sin u$ & $0$ & $s^*$ & always &
unstable for \\ 
&    &   &   &   &  &    & $0\leq \gamma<2, 0<u<\pi,$
\\[0.2cm]  
  &   &   &   &   &   &   & $s^*<\sqrt{6} \csc (u),
f'\left(s^*\right)<0$ or \\
 &    &   &   &   &  &    & $0\leq \gamma<2,\pi<u< 2\pi,$
\\  
  &   &   &   &   &   &   & $s^*>\sqrt{6} \csc
(u),
f'\left(s^*\right)>0$  \\[0.2cm]
 &    &   &   &   &  &    &   saddle otherwise\\
 \hline\hline
 \end{tabular}} 
\end{center}
\caption{\label{critKS}The critical points and curves of critical points of
the
system \eqref{KSQ}-\eqref{KSs} of Kantowski-Sachs
scalar-field cosmology. 
We use the notation $\epsilon=\pm1$,  where $\epsilon=+1$ corresponds to
expanding universe and $\epsilon=-1$   to contracting
one, with the stability conditions outside  parentheses
corresponding to $\epsilon=+1$ while those inside parentheses
  to $\epsilon=-1$. We use the notation $s^*$ for
the real
values of $s$ satisfying $f(s^*)=0$, and $s_c$ for an a wide range of real
value.  Furthermore, the variable
$u$ varies in $[0,2\pi]$. Finally, note that the points $P_7^\epsilon$ and
$P_8^\epsilon$ are special points of the curves $C_\epsilon(0)$, and are
given separately for clarity.}
\end{table*}

There are several invariant sets, that is areas of the phase-space
that evolve to themselves under the dynamics, for the dynamical
system \eqref{KSQ}-\eqref{KSs}. Firstly, we have the invariant
sets $Q=\pm1$, corresponding to ${\cal K}=0$,
where in particular $Q=1$ corresponds to expanding universe,
whereas $Q=-1$ corresponds to contracting one.
Furthermore, there exist invariant sets corresponding to specific
subclasses of the general scenario, such is the  set $x^2+y^2+\Sigma^2=1$,
that is $z=0$, which  corresponds to absence of standard matter,
and which includes the isotropic invariant set $x^2+y^2=1,\,
\Sigma=0$, or the set $y=0$ which corresponds to potential
absence, or the set $x=y=0$ which corresponds to scalar-field absence.

The scenario of Kantowski-Sachs scalar-field cosmology (the system
\eqref{KSQ}-\eqref{KSs}) admits twelve isolated critical points  (six
corresponding to expanding universe and six corresponding  to
contracting one) and  ten
curves of critical points (five corresponding to expanding universe and five
 corresponding  to
contracting one), which are displayed in Table \ref{critKS} along
with their existence and stability conditions \footnote{Note that for
particular cases of $f(0)$ and $f'(0)$ some of these points become
non-hyperbolic,  and therefore in order to extract their stability
properties a center manifold analysis \cite{wiggins} is necessary. However,
since such an investigation in the general case lies beyond the scope of the
present work, we prefer to perform it straightaway in the specific
applications of the following sections, whenever it is necessary.}. We use
the notation
$\epsilon=\pm1$, that is $\epsilon=+1$  corresponds to
expanding universe, while $\epsilon=-1$ corresponds to contracting
one. The details of the analysis and the calculation of the various
eigenvalues of the $5\times5$
 perturbation matrix ${\bf {\Xi}}$   are
presented in  \ref{Eigenvalues1}. 
Moreover,
for each critical point we calculate the values of 
the various density parameters, the   deceleration
parameter 
and the dark-energy and total equation-of-state parameters, given by
(\ref{OmegasKSaux}), and we summarize the results in Table
\ref{critKS2}.

Once again we stress that all these results hold for a wide range of
potentials, that is why there is a large variety of critical points and
curves of critical points. Interestingly enough, as we are going to see in
detail in the
specific applications of the following sections, for some of the usual
potentials of the literature the majority of these stable points disappear or
are not stable anymore, and that is why they have not been obtained in the
specific-potential literature
\cite{Burd:1988ss,Byland:1998gx,Copeland:1997et,Coley:2003mj}. This
feature may be an indication that some of the simple potentials of the
literature, such is the exponential and the power-law one, might restrict the
dynamics in scalar-field cosmology.  

\begin{table*}[ht]
\begin{center}
\resizebox{\columnwidth}{!}{
\begin{tabular}{cccccccc}
\hline \hline
  Name & $\Omega_k$ &$\Omega_m$ & $\Omega_{DE}$ & $\Omega_\sigma$  & $q$ &
$w_{DE}$ & $w_{tot}$ \\
\hline\hline
$P_1^\epsilon$ & 0&0&0&1&2&Arbitrary& 1\\\vspace{0.2cm}
$P_2^\epsilon$ & 0&0&0&1&2&Arbitrary& 1\\\vspace{0.2cm}
 $P_3^\epsilon$ & $\frac{3}{4} (2-\gamma) (3 \gamma -2)$ & $3 (1-\gamma)$
&0& $\frac{1}{4} (3 \gamma -2)^2$ & $\frac{3 \gamma }{2}-1$ &Arbitrary
&$\gamma
   -1$ \\
$P_4^\epsilon$ & 0&1&0&0& $\frac{3 \gamma }{2}-1$ &Arbitrary& $\gamma
-1$
\\\vspace{0.2cm}\\
$P_5^\epsilon$& 0&0&1&0&$-1$&$-1$ & $-1$
\\\vspace{0.2cm}\\
$P_6^\epsilon$ & $-3$&0&3&1&$-1$& $-1$ &$-1$   \\\vspace{0.2cm}\\
$P_7^\epsilon$ & 0&0&1&0&2&1&1
\\\vspace{0.2cm}\\
$P_8^\epsilon$ & 0&0&1&0&2&1&1
\\ \vspace{0.2cm}\\
$P_9^\epsilon(s^*)$& 0&$1-\frac{3 \gamma}{\left(s^*\right)^2}$ &$\frac{3
\gamma
}{\left(s^*\right)^2}$ &0& $\frac{3 \gamma }{2}-1$& $\gamma -1$ & $\gamma
-1$\\\vspace{0.2cm}\\
$P_{10}^\epsilon(s^*)$ & $\frac{3 \left[\left(s^*\right)^4-4\right]}{4
\left[\left(s^*\right)^2+1\right]^2}$&0&$\frac{3}{\left(s^*\right)^2+1}
$&$\frac{\left[\left(s^*\right)^2-2\right]^2}{4
   \left[\left(s^*\right)^2+1\right]^2}$&$\frac{\left(s^*\right)^2-2}{2
\left[\left(s^*\right)^2+1\right]}$&$-\frac{1}{\left(s^*\right)^2+1}$&
$-\frac{1}{\left(s^*\right)^2+1}$
\\\vspace{0.2cm}\\
$P_{11}^\epsilon(s^*)$ & 0&0&1&0&$ 
 \frac{\left(s^*\right)^2}{2}-1$ &   $ 
 \frac{\left(s^*\right)^2}{3}-1$   
& $ 
 \frac{\left(s^*\right)^2}{3}-1$   
\\\vspace{0.2cm}\\
$P_{12}^\epsilon(s^*)$ & 0&$1-\frac{3 \gamma }{\left(s^*\right)^2}$&$\frac{3
\gamma }{\left(s^*\right)^2}$&0&$\frac{3 \gamma }{2}-1$&$\gamma -1$&$\gamma
-1$
\\\vspace{0.2cm}
$C_\epsilon(s^*)$& 0&0&$\sin ^2 u$&$\cos ^2 u$&2&1&1\\
 \hline\hline
 \end{tabular}}
\end{center}
\caption{\label{critKS2}
The critical points and curves of critical points of the
system \eqref{KSQ}-\eqref{KSs} of Kantowski-Sachs
scalar-field cosmology and the
corresponding values of the 
basic observables, namely the curvature density
parameter $\Omega_k$, the matter density parameter $\Omega_m$, the
dark-energy density parameter $\Omega_{DE}$, the shear density
parameter $\Omega_\sigma$, the deceleration parameter $q$, the  
dark-energy equation-of-state parameter $w_{DE}$ and the total
equation-of-state parameter $w_{tot}$, calculated using
(\ref{OmegasKSaux}).
The notation is as in Table    \ref{critKS}. }
\end{table*}

Amongst the above various critical points, we are interested those which
correspond to an expanding universe (that is with
$\epsilon=+1$), which moreover are stable and thus they can be the
late-time state of the universe. In particular:

\begin{itemize}

 \item  Point $P_5^+$, which
is asymptotically stable for potentials having $f(0)>0$, corresponds to the
de Sitter ($w_{tot}=-1$), isotropic ($\Omega_\sigma=0$), accelerating ($q<0$)
universe, which is dark-energy
dominated  ($\Omega_{DE}=1$), with the dark energy behaving like a
cosmological constant ($w_{DE}=-1$).

\item Point
$P_{11}^+(s^*)$ corresponds to an isotropic, dark-energy dominated
universe, with the dark-energy equation-of-state parameter lying in the
quintessence regime, which can be accelerating or not according to the
potential parameters. In the case of exponential potential this point
becomes the most important one, since   it is both stable and compatible
with observations   \cite{Copeland:1997et}.  

\item Point $P_{12}^+(s^*)$
corresponds to an isotropic universe with $0<\Omega_{DE}<1$, that is it can
alleviate
the coincidence problem since dark-energy and dark-matter density
parameters can be of the same order.  However, it has the disadvantage that
for the usual case of dust matter ($\gamma=1$) it is not accelerating and
moreover it leads to
$w_{DE}=0$, which are not favored by
observations.

 \end{itemize}

Finally, note that in the present case the regions of expanding
($\epsilon=+1$) and contracting ($\epsilon=-1$) universe are not
disconnected, and thus theoretically one could have heteroclinic orbits from
one to the other region, which is the  realization of a cosmological
bounce. However, as we prove in the end of this section, this is not
possible as long as we have a real minimally-coupled scalar field and a
matter sector satisfying the null energy condition.

\subsection{LRS Bianchi III and Bianchi I scalar-field cosmology}
\label{BIIIphase_space}

{{
Contrary to the Kantowski-Sachs case,  where Hubble normalized variables are
not compact and thus a separate analysis is needed, in  
LRS Bianchi III and Bianchi I geometries one can perform the
dynamical analysis in a unified way. The corresponding cosmological
equations are
\eqref{KSsfpfGauss},\eqref{KSsfpfpropK},\eqref{KSsfpfevolanisotropies},
\eqref{KSsfpfRaych1},\eqref{KSsfpfevolmatter},\eqref{KSsfpfKG} for the
choices $k=-1$ and $k=0$, respectively. 
In order to transform these cosmological system  in its
autonomous form we introduce the auxiliary variables:
\begin{equation}
x=\frac{\dot\phi}{\sqrt{6} H},\; y=\frac{\sqrt{V}}{\sqrt{3}H},\;
 z=\frac{\rho_m}{3H^2},\; \Sigma=\frac{\sp}{H},\; \Omega_k=-\frac{k
K}{3H^2},\label{BIIIvars}
\end{equation}
using also the   variable $\tau$ defined through
$ {d}\tau=|H|{d} t\equiv \epsilon H {d} t$, where $\epsilon=\text{sgn}(H)$ 
(primes will denote derivatives with
respect to $\tau$). Additionally, we use the two variables $s$ and $f$
defined in (\ref{sdef}), (\ref{fdef}), in order to handle the unspecified 
potential $V(\phi)$.  

We mention here that for the Bianchi I case $\Omega_k$ becomes automatically
zero (since $k=0$), and thus examining both LRS Bianchi III and LRS Bianchi I
in a unified way, Bianchi I corresponds to the invariant set
$\Omega_k=0$.\footnote{
An alternative way to obtain the LRS Bianchi I case is from the
corresponding Kantowski-Sachs one, setting
$Q=\pm 1$ and $D=|H|$, since $K$ does not appear explicitly in the LRS
Bianchi I cosmological equations (note however
that this does not mean that $K=0$).} However, note that in Bianchi III one
can also have some critical points with $\Omega_k=0$ as expected, that also
belong to the Bianchi I boundary. The difference
is that in the latter case the stability will be in principle different,
since the system can evolve in the $\Omega_k$-direction, which is not the
case in Bianchi I.
%The dynamical behavior in this invariant subset will be discussed in the
%paragraph \ref{BIphase_space}.  

From the definitions (\ref{BIIIvars}) we deduce that
$-1\leq  x\leq 1,$ $-1\leq  y\leq
1,$  $ z\geq 0$  and $\Omega_k\geq 0$,   while $s$ is un-constrained.
  Moreover, in terms of the
auxiliary variables   the Gauss constraint \eqref{KSsfpfGauss}
becomes  $x^2+y^2+ z+\Sigma^2+\Omega_k=1$, which allows us to reduce the
dimension of the phase space, eliminating one dynamical variable, for
instance $z$.

%{\bf  Similarly to Bianchi I analysis, we mention that in the special case
%where $H=0$ the auxiliary variables (\ref{BIIIvars}) may diverge, and thus
%an analysis at the infinity,  introducing Poicar\'e compact variables  
%\cite{Byland:1998gx,Xu:2012jf}, would be needed for completeness. For
%simplicity we omit such analysis in the present investigation.}

Finally, in
terms of the
auxiliary
variables the various density parameters, the deceleration
parameter and the dark-energy and total
equation-of-state
parameters, defined in
(\ref{Omegas1})-(\ref{wDtot}), read:
\begin{eqnarray}
 %&& -\Omega_k=x^2+y^2+z+\Sigma ^2-1,\nonumber\\
 &&  \Omega_{m}=1-x^2-y^2-\Sigma^2-\Omega_k,\nonumber\\
 &&\Omega_{DE}\equiv\Omega_{\phi}=x^2+y^2,\nonumber\\
&&\Omega_\sigma=\Sigma ^2,\nonumber\\
&&q=\frac{3}{2}(2-\gamma)\left(x^2+\Sigma^2\right)-\frac{3\gamma}{2}y^2+\frac
{1}{2}(3\gamma-2)\left(1-\Omega_k\right), \nonumber\\
 &&w_{DE}=-1+\frac{2
   x^2}{x^2+y^2},\nonumber\\
    &&w_{tot}=-1+\gamma +\frac{(2-\gamma ) x^2-\gamma  y^2}{1-\Sigma
^2-\Omega_k}.
\label{OmegasBIIIaux}
\end{eqnarray}

 In summary, using the dimensionless auxiliary variables
\eqref{BIIIvars}, along with the  constraint equation, we result to the
five-dimensional dynamical system:
\begin{align}
& \Sigma'= \epsilon\left\{\left(3-\frac{3 \gamma }{2}\right) \Sigma
^3-\frac{1}{2} \Sigma  \left[3 (\gamma -2) x^2+3 \gamma 
\left(y^2+\Omega_k-1\right)-2 (\Omega_k-3)\right]+\Omega_k\right\},
\label{BIIISigma}\\
& x'=\epsilon\left\{\frac{3}{2}\left(2-\gamma\right) x^3+\frac{1}{2} x
\left[2 \left(3 \Sigma ^2+\Omega_k-3\right)-3 \gamma  \left(\Sigma
^2+y^2+\Omega_k-1\right)\right]+\sqrt{\frac{3}{2}} s y^2\right\},
\label{BIIIx}\\
& y'=\epsilon\left\{-\frac{3 \gamma 
   y^3}{2}+\frac{1}{2} y \left[-3 \gamma  \left(\Sigma
^2+\Omega_k-1\right)-\sqrt{6} s x+2 \left(3 \Sigma ^2+\Omega_k\right)-3
(\gamma -2) x^2\right]\right\},\label{BIIIy}\\
&  \Omega_k'=\epsilon\left\{(2-3 \gamma ) \Omega_k^2+\Omega_k \left[-3
(\gamma -2) \Sigma ^2+3 \gamma -2 \Sigma +(6-3 \gamma ) x^2-3 \gamma 
y^2-2\right]\right\},\label{BIIIz}\\
& s'=-\epsilon\sqrt{6} x f(s),
\label{BIIIs}
\end{align}
defined on the phase space 
\begin{align}
& \{(\Sigma,x,y,\Omega_k,s)\in\mathbb{R}^5:-1\leq
\Sigma\leq1,-1\leq x\leq1, 0\leq y\leq1, 0\leq \Omega_k\leq1, \nonumber \\ &
\ \ \ 
\ \ \ \ \ \ \ \ \ \  0\leq  x^2+y^2+\Omega_k+\Sigma^2\leq 1\}.
\end{align}
}}
As before, $\epsilon=+1$ corresponds to an expanding universe
while    $\epsilon=-1$   to a contracting one. However, we mention that
in the present case, and contrary to the previous Kantowski-Sachs one,
these two regions, that is
$y<0$ and $y>0$,   are not connected, since the sign of $y$ is an invariant,
as can be deduced from (\ref{BIIIy}). Therefore, a transition from one to
the other is impossible, and thus LRS Bianchi III and LRS Bianchi I
scalar-field cosmologies do
not allow for a cosmological bounce. This is verified by the analysis of
subsection  \ref{Sect:bounce}, where we prove that   a bounce is not
possible as long as we have a real minimally-coupled scalar field and a
matter sector satisfying the null energy condition.

{{Furthermore, observe that the system
\eqref{BIIISigma}-\eqref{BIIIs} is
form-invariant under the change 
  \begin{equation}
  \label{(78)}\{\epsilon, \Sigma, x, y, \Omega_k,
s\} \rightarrow \{-\epsilon, -\Sigma, -x, -y, \Omega_k, s\},
\end{equation} 
and thus  if a point is unstable in the expanding branch
then its symmetrical partner in the contracting branch will be stable for
the same conditions, and viceversa. }} Therefore,
it is sufficient to discuss the behavior of one half of the phase space, and
the dynamics in the other half will be obtained via the transformation
\eqref{(78)}. Hence, we restrict our analysis to the positive
$\epsilon=+1$ (expanding) branch.

{ There are several invariant sets, for the system
\eqref{BIIISigma}-\eqref{BIIIs} of LRS Bianchi III and I expanding 
($\epsilon=+1$)
scalar-field cosmology.  Amongst them there exists the invariant set
$\Omega_k=0$
(corresponding to $k=0$ since $\ey\neq 0$), which is the boundary
corresponding to LRS Bianchi I scalar-field cosmology. This
invariant set contains the isotropic invariant set
$\Omega_k=0$, $\Sigma=0$.} Finally, there exist the invariant
sets $y=0$ (potential absence), and the
invariant set $x=y=0$ (scalar-field absence).

 \begin{table*}[ht]
\begin{center}
\resizebox{\columnwidth}{!}{
\begin{tabular}{ccccccccc}
\hline \hline
  Name & $\Sigma$ & $x$ & $y$ & $\Omega_k$ & $s$ & Existence & BIII & BI  \\
\hline\hline
  $R_1^+$ & $ 1$ & $0$ & $0$ & $0$ & $s_c$ & always & \checkmark & \checkmark
 \\\vspace{0.2cm}
 $R_1^-$ & $- 1$ & $0$ & $0$ & $0$ & $s_c$ & always & \checkmark & \checkmark
 \\\vspace{0.2cm}
  $R_2$ & $\frac{1}{2}$ & $0$ & $0$ & $\frac{3}{4}$ & $s_c$ & always & \checkmark & \xmark
\\\vspace{0.2cm}
  $R_3$ & $0$ & $0$ & $0$ & $0$ & $s_c$ & always & \checkmark & \checkmark \\\vspace{0.2cm}
  $R_4$ & $-1+\frac{3\gamma}{2}$ & $0$ & $0$ &
$\frac{3}{4}(2-\gamma)(3\gamma-2)$ & $s_c$ &
$\frac{2}{3}\leq\gamma\leq 1$ & \checkmark & \xmark \\\vspace{0.2cm}
  $R_5$ & $0$ & $0$ & $1$ & $0$ & $0$ & always & \checkmark & \checkmark \\\vspace{0.2cm}
  $R_6(s^*)$ & $0$ & $\frac{\sqrt{3} \gamma }{\sqrt{2}s^*}$ &
$\sqrt{\frac{3}{2}} \sqrt{\frac{(2-\gamma ) \gamma }{\left({s^*}\right)^2}}$
& $0$ & $s^*$ & $\left({s^*}\right)^2\geq
3\gamma, s^*\neq 0$ & \checkmark & \checkmark \\\vspace{0.2cm}
  $R_7(s^*)$ & $\frac{3 \gamma }{2}-1$ & $\frac{\sqrt{3} \gamma
}{\sqrt{2}s^*}$ & $\sqrt{\frac{3}{2}} \sqrt{\frac{(2-\gamma) \gamma
   }{\left({s^*}\right)^2}}$& $\frac{3}{4}(2-\gamma)(3\gamma-2)$ & $s^*$ &
$\frac{2}{3}\leq \gamma<1,\, \left({s^*}\right)^2\geq
\frac{\gamma}{1-\gamma}$ & \checkmark & \xmark
\\\vspace{0.2cm}
     $R_8(s^*)$ & $0$ & $\frac{s^*}{\sqrt{6}}$ &
$\sqrt{1-\frac{\left({s^*}\right)^2}{6}}$&
$0$ & $s^*$ & $\left({s^*}\right)^2<6, s^*\neq 0$ & \checkmark & \checkmark  \\\vspace{0.2cm}
$R_9^+(s^*)$ & $0$ & $ 1$ & $0$& $0$ & $s^*$ & always & \checkmark & \checkmark \\\vspace{0.2cm}
$R_9^-(s^*)$ & $0$ & $- 1$ & $0$& $0$ & $s^*$ & always & \checkmark & \checkmark\\\vspace{0.2cm}
$R_{10}(s^*)$ & $\frac{\left[(s^*)^2-2\right]}{2\left[(s^*)^2+1\right]}$
& $\frac{\sqrt{\frac{3}{2}}s^*}{(s^*)^2+1}$ &
$\frac{\sqrt{\frac{3}{2}}
\sqrt{\left(s^*\right)^2+2}}{\left(s^*\right)^2+1}$&
$\frac{3 \left[\left(s^*\right)^4-4\right]}{4
\left[\left(s^*\right)^2+1\right]^2}$ & $s^*$ & $\left({s^*}\right)^2\geq 2$ & \checkmark & \xmark
\\\vspace{0.2cm}
$C(s^*)$ & $\cos u$ & $\sin u$ & $0$ & $0$ & $s^*$ & always & \checkmark & \checkmark  \\
  \hline
\end{tabular}}
\end{center}
\caption{\label{critBIII}   
The critical points and curves of critical points of the
system \eqref{BIIISigma}-\eqref{BIIIs} of LRS Bianchi III and Bianchi I
scalar-field cosmologies, in the expanding universe sub-space ($\epsilon=+1$). 
We  use the notation $s^*$ for
the real
values of $s$ satisfying $f(s^*)=0$, and $s_c$ for an arbitrary of real
value.  Finally, the variable
$u$ varies in $[0,2\pi]$. }
\end{table*}

%%%%%%%%%NEW TABLES%%%%%%%

 \begin{table*}[!]
\begin{center}
\resizebox{\columnwidth}{!}
{
\begin{tabular}{ccc}
\hline \hline
  Name  & Stability of LRS BIII models & Stability restricted to LRS BI boundary\\
\hline\hline
  $R_1^+$ 
& unstable & unstable\\\vspace{0.2cm}
 $R_1^-$ 
& unstable & unstable\\\vspace{0.2cm}
  $R_2$ & saddle & \xmark
\\\vspace{0.2cm}
  $R_3$ & saddle & saddle \\\vspace{0.2cm}
  $R_4$ & saddle & \xmark \\\vspace{0.2cm}
  $R_5$ & stable for $f(0) > 0$ or  
saddle otherwise & same as for LRS BIII \\\vspace{0.2cm}
  $R_6(s^*)$ & stable for & stable for 
 \\
	  &  $0<\gamma<\frac{2}{3},\,
s^*<-\sqrt{3\gamma}$ and
$f'(s^*)<0,$  & $0<\gamma<2,\,
s^*<-\sqrt{3\gamma}$ and
$f'(s^*)<0,$
\\\vspace{0.2cm}
 &   or  $0<\gamma<\frac{2}{3},\,
s^*>\sqrt{3\gamma}$
and $f'(s^*)>0;$ &   or  $0<\gamma<2,\,
s^*>\sqrt{3\gamma}$
and $f'(s^*)>0;$
\\\vspace{0.2cm}
    &   saddle otherwise &   saddle otherwise
\\\vspace{0.2cm}
  $R_7(s^*)$ & stable for & \xmark \\
 &   $\frac{2}{3}<\gamma
<1,s^*<-\sqrt{\frac{\gamma }{1-\gamma}}, f'\left(s^*\right)<0$    
\\\vspace{0.2cm}
  &   or $\frac{2}{3}<\gamma <1,
   s^*>\sqrt{\frac{\gamma }{1-\gamma}}, f'\left(s^*\right)>0$
\\\vspace{0.2cm}
   &   saddle otherwise
\\\vspace{0.2cm}
     $R_8(s^*)$  & stable for & stable for  \\
    &   $0<\gamma\leq \frac{2}{3},
-\sqrt{3\gamma}<s^*<0, f'(s^*)<0$   &   $0<\gamma<2,
-\sqrt{3\gamma}<s^*<0, f'(s^*)<0$  
\\\vspace{0.2cm}
  &  or $0<\gamma\leq \frac{2}{3},
0<s^*<\sqrt{3\gamma}, f'(s^*)>0$ &  or $0<\gamma<2,
0<s^*<\sqrt{3\gamma}, f'(s^*)>0;$
\\\vspace{0.2cm}
  &   $\frac{2}{3}<\gamma\leq 2, -\sqrt{2}<s^*<0,
f'(s^*)<0$   & saddle otherwise
\\\vspace{0.2cm}
  & or  $\frac{2}{3}<\gamma\leq 2, 0<s^*<\sqrt{2},
f'(s^*)>0$
\\\vspace{0.2cm}
  &   saddle otherwise  \\\vspace{0.2cm}
      $R_9^+(s^*)$ & unstable for 
$0\leq \gamma<2$, $s^*<0$,  
$f'\left(s^*\right)<0$ & same as for LRS BIII \\\vspace{0.2cm}
 &   saddle otherwise  \\\vspace{0.2cm}
     $R_9^-(s^*)$ & unstable for 
$0\leq \gamma<2$, $s^*>0$,  
$f'\left(s^*\right)>0$ & same as for LRS BIII \\\vspace{0.2cm}
&   saddle otherwise 
\\\vspace{0.2cm}
     $R_{10}(s^*)$ & stable for & \xmark\\
&   $\frac{2}{3}<\gamma<1,
-\sqrt{\frac{\gamma}{1-\gamma}}<s^*<-\sqrt{2}, f'(s^*)<0$   
\\\vspace{0.2cm}
 &  or $1\leq \gamma\leq 2,
s^*<-\sqrt{2}, f'(s^*)<0$
\\\vspace{0.2cm}
 &   $\frac{2}{3}<\gamma<1,
\sqrt{2}<s^*<\sqrt{\frac{\gamma}{1-\gamma}},
f'(s^*)>0$   
\\\vspace{0.2cm}
 & or  $1\leq \gamma\leq 2, s^*>\sqrt{2},
f'(s^*)>0$
\\\vspace{0.2cm}
$C(s^*)$ & unstable for & same as for LRS BIII
\\\vspace{0.2cm}
&
  $0\leq \gamma<2, 0<u<\pi, s^*<\sqrt{6} \csc (u),
f'\left(s^*\right)<0$ or  \\\vspace{0.2cm}
&
 $0\leq \gamma<2,\pi<u< 2\pi, s^*>\sqrt{6} \csc (u),
f'\left(s^*\right)>0$  \\\vspace{0.2cm}
&
saddle otherwise  \\
  \hline
\end{tabular}}
\end{center}
\caption{\label{critBIIIb}   
Stability conditions for the critical points and curves of critical points of the
system \eqref{BIIISigma}-\eqref{BIIIs} of both LRS Bianchi III and LRS Bianchi I
scalar-field cosmologies, in the expanding universe sub-space ($\epsilon=+1$). 
We use the same notations as in table \ref{critBIII}.}
\end{table*}
 
%%%%%%%%%%%%%%%%%%%%%%%%%%
%%%%%%%%%%%%%%%%%%%%%%%%%%%

{ The scenario of LRS  Bianchi III scalar-field cosmology (the system
\eqref{BIIISigma}-\eqref{BIIIs}), in the expanding universe sub-space, admits
six isolated critical points and seven curves of critical points, which are
displayed in Table \ref{critBIII} along with their existence conditions.
Amongst them, the points satisfying $\Omega_k=0$ exists for LRS  Bianchi I
too. However, as we mentioned above, the stability of the points with
$\Omega_k=0$ will be different in the two geometry classes. In particular,
the stability analysis of LRS Bianchi I does not require the examination of
the
stability along the $\Omega_k$-axis, while in LRS  Bianchi III this is
necessary since the system can move along the $\Omega_k$-direction too.
 % However, if we
%analyze the stability of the same point as a LRS BIII model it is required
%the examination along the $\Omega_k$ extra direction and in this cases new
%bifurcation values and new stability conditions appear. 
%
%The resulting conditions provide information about the stability against
%curvature perturbations of some specific LRS BI models when viewed as an
%invariant set of the more general Bianchi III geometries. 
In Table
\ref{critBIIIb} we summarize the stability conditions for LRS  Bianchi III
and LRS  Bianchi I cases.
} The details of the
analysis and the calculation of the various eigenvalues of the $5\times5$
perturbation matrix ${\bf {\Xi}}$ are presented in 
\ref{Eigenvalues3}. \footnote{Observe that for LRS Bianchi I models the
perturbation matrix is $4\times 4$, since the direction $\Omega_k$ is absent.
The eigenvalues will be the same as for the $5\times 5$
matrix, apart from the first one that corresponds to the extra
$\Omega_k$-direction.}

Finally, for each critical point we calculate the values of 
the various density parameters, the deceleration
parameter 
and the dark-energy and total
equation-of-state
parameters, given by
(\ref{OmegasBIIIaux}), and we summarize the results in Table
\ref{critBIII2}. We mention that the above points and curves correspond only
to the half phase-space of expanding solutions. Thus, the whole phase-space
admits also their symmetric partners corresponding to contractions, which
coordinates and observables are given by the expanding ones under the
transformation (\ref{(78)}).

\begin{table*}[!]
\begin{center}\resizebox{\columnwidth}{!}{
\begin{tabular}{cccccccc}
\hline \hline
  Name & $\Omega_k$ &$\Omega_m$ & $\Omega_{DE}$ & $\Omega_\sigma$  & $q$ &
$w_{DE}$ & $w_{tot}$\\
\hline\hline
$R_1^+$  & 0&0&0&1&2& Arbitrary & $1$  \\\vspace{0.2cm}
$R_1^-$  & 0&0&0&1&2& Arbitrary& $1$  \\\vspace{0.2cm}
$R_2$ & $\frac{3}{4}$&0&0&$\frac{1}{4}$&$\frac{1}{2}$& Arbitrary &$0$
\\\vspace{0.2cm}
 $R_3$ &0&1&0&0&   $\frac{3 \gamma
   }{2}-1$   & Arbitrary& $\gamma -1$  \\
$R_4$ & $\frac{3}{4}(2-\gamma)(3\gamma-2)$&$3-3 \gamma$&0&$\left(1-\frac{3
\gamma }{2}\right)^2$&   $\frac{3 \gamma
   }{2}-1$       &Arbitrary&$\gamma -1$  \\\vspace{0.2cm}\\
$R_5$& 0&0&1&0&$-1$&$-1$&$-1$
\\\vspace{0.2cm}\\
$R_6(s^*)$ & 0 & $1-\frac{3 \gamma }{\left(s^*\right)^2}$&$\frac{3 \gamma
}{\left(s^*\right)^2}$&0&$\frac{3 \gamma }{2}-1$&$\gamma -1$&$\gamma
   -1$\\\vspace{0.2cm}\\
$R_7(s^*)$ & $\frac{3}{4}(2-\gamma)(3\gamma-2)$ &$-3 \gamma -\frac{3 \gamma
}{\left(s^*\right)^2}+3$&$\frac{3 \gamma
}{\left(s^*\right)^2}$&$\left(1-\frac{3 \gamma }{2}\right)^2$&$\frac{3
   \gamma }{2}-1$&$\gamma -1$&$\gamma -1$
\\\vspace{0.2cm}\\
$R_8(s^*)$ &0&0&1&0&    $ 
 \frac{\left(s^*\right)^2}{2}-1$  &  $ 
 \frac{\left(s^*\right)^2}{3}-1$    &$ 
 \frac{\left(s^*\right)^2}{3}-1$ 
\\\vspace{0.2cm}\\
$R_9^+(s^*)$&
 0 & 0&1&0&2&1&1 \\ 
\vspace{0.2cm}\\
$R_9^-(s^*)$&
 0 & 0&1&0&2&1&1 \\ 
\vspace{0.2cm}\\
$R_{10}(s^*)$&$\frac{3 \left[\left(s^*\right)^4-4\right]}{4
\left[\left(s^*\right)^2+1\right]^2}$&$0$&$\frac{3}{\left(s^*\right)^2+1}
$&$\frac{\left[\left(s^*\right)^2-2\right]^2}{4
   \left[\left(s^*\right)^2+1\right]^2}$&$\frac{\left(s^*\right)^2-2}{2
\left[\left(s^*\right)^2+1\right]}$&$-\frac{1}{\left(s^*\right)^2+1}$&$-\frac
{1}{\left(s^*\right)^2+1}$
\vspace{0.2cm}\\
$C(s^*)$& 0 & 0&$\sin ^2 u$&$\cos ^2 u$&2&1&1\\
 \hline\hline
 \end{tabular}} 
\end{center}
\caption {\label{critBIII2} 
The critical points and curves of critical points of the
system \eqref{BIIISigma}-\eqref{BIIIs} of both LRS Bianchi III and Bianchi I
scalar-field cosmologies, in the expanding universe sub-space ($\epsilon=+1$), 
and the
corresponding values of the 
basic observables, namely the curvature density
parameter $\Omega_k$ (which is zero for BI models), the matter density parameter $\Omega_m$, the
dark-energy density parameter $\Omega_{DE}$, the shear density
parameter $\Omega_\sigma$, the deceleration parameter $q$, the  
dark-energy equation-of-state parameter $w_{DE}$ and the total
equation-of-state parameter $w_{tot}$, calculated using
(\ref{OmegasBIIIaux}). The notation is as in Table \ref{critBIII}.}
\end{table*}
 
Lastly, we stress that all these results hold for a wide range of potentials,
that
is why there is a large variety of critical points and curves of critical
points. Similarly to the previous  Kantowski-Sachs case,
for
some of the usual potentials the majority of these stable points disappear or
are not stable anymore, which may be an indication that some of the simple
potentials of the literature, such is the exponential and the power-law one,
might restrict the dynamics in scalar-field cosmology.

The physically interesting critical points are those which correspond to  
expansion, which moreover are stable
and thus they can be the late-time state of the universe. In particular:

\begin{itemize}

 \item  Point $R_5$, which is asymptotically stable for potentials having
$f(0)>0$, corresponds to the de Sitter ($w_{tot}=-1$), isotropic
($\Omega_\sigma=0$), accelerating ($q<0$) universe, which is dark-energy
dominated  ($\Omega_{DE}=1$), with the dark energy behaving like a
cosmological constant ($w_{DE}=-1$). {{This point exists for both
LRS Bianchi III and Bianchi I geometries.}}

\item Point $R_6(s^*)$ corresponds to an isotropic universe with
$0<\Omega_{DE}<1$, that is it can alleviate the coincidence problem since
dark-energy and dark-matter density parameters can be of the same order. 
However, it has the disadvantage that for the usual case of dust matter
($\gamma=1$) it is not accelerating and moreover it leads to $w_{DE}=0$,
which are not favored by observations.  {{This point exists for both
LRS Bianchi III and Bianchi I geometries, however the stability intervals in
the two cases are different.}}

\item Point $R_7(s^*)$ corresponds to an anisotropic universe, with
$0<\Omega_{DE}<1$, $0<\Omega_m<1$ and
non-vanishing $\Omega_k$. It is not accelerating and  for usual dust
matter it leads to $w_{DE}=0$. Thus,
this point is not favored by observations. However, it is still very
interesting that this point, although stable, maintains a non-zero
anisotropy. We discuss on these issues in the subsection \ref{Sect:bounce}. 
{{This point exists only for
LRS Bianchi III geometry.}}

\item Point $R_8(s^*)$
corresponds to an isotropic, dark-energy dominated universe, with the
dark-energy equation-of-state parameter lying in the
quintessence regime, which can be accelerating or not according to the
potential parameters. In the case of exponential potential this point
becomes the most important one, since   it is both stable and compatible
with observations \cite{Copeland:1997et}. {{This point exists for both
LRS Bianchi III and Bianchi I geometries, however the stability intervals in
the two cases are different.}}

\item { Point $R_{10}(s^*)$ corresponds to an anisotropic universe, with
$0<\Omega_{DE}<1$, $0<\Omega_\sigma<1$, non-vanishing $\Omega_k$ and
$\Omega_m=0$, with the
dark-energy equation-of-state parameter lying in the
quintessence regime. It is not accelerating and thus
this point is not favored by observations. However, 
interestingly enough this point, although it can be stable, maintains a
non-zero
anisotropy. We discuss on these issues in the subsection \ref{Sect:bounce}.
This point exists only for
LRS Bianchi III geometry.
}
 \end{itemize}

\subsection{Late-time isotropization and  Bounce behavior}
\label{Sect:bounce}
 
 Having performed the dynamical analysis for the three geometrical classes,
we desire to address two important issues of physical significance, namely
the late-time isotropization and the bounce behavior.

\subsubsection{Late-time isotropization.}  

The criterion of late-time  isotropization in an expanding universe ($H>0$)
is the vanishing of the shear $\sigma_+$
\cite{Mendes:1990eb}, or alternatively we can use the stronger condition
\cite{Burd:1988ss,Byland:1998gx,Collins:1972tf}:
\begin{equation} 
\frac{\sigma_+}{H}\rightarrow 0 \qquad \text{as}\quad t\rightarrow\infty.
\end{equation}
%Since we are interested in expanding solutions we suppose that $H$ is
%positive or equal to zero at some time $t_0.$ 

For LRS Bianchi  I and III geometries, with a  scalar field with a positive
and convex potential, which
is expanding at a given time $t=t_0$, that is $H(t_0)>0$, it is
well-known that
\cite{Byland:1998gx}:
\begin{enumerate} 
\item[(i)] $ H(t) \geq 0, \dot H(t)\leq 0$ for all $t\geq t_0$,
\item[(ii)] $\sigma_+(t), \dot\phi(t), {}^{(3)}\! R(t)\rightarrow 0$ for
$t\rightarrow \infty$, 
\item[(iii)] $H\rightarrow \sqrt{\frac{V_0}{3}}$, where $V_0 $ is the minimum
of the potential and in particular $H\rightarrow 0$ for an exponential
potential. 
\end{enumerate}
These results can be also straightforwardly proven in the case of an
additional barotropic fluid satisfying the strong  energy conditions namely 
$\rho_m+p_m>0$ and $\rho_m+3 p_m>0$, since at
late times $\rho_m\rightarrow 0$ and then the scalar field dominates.
However, the above  theorem does not give an answer to isotropization
for the cases where the potential minimum is zero, or in the case
of exponential potentials, since $\sigma_+$ and $H$
vanish simultaneously, thus the rate $\sigma_+/H$ cannot be calculated
 {\it{a priori}} from the field equations. Furthermore, for more general
potentials, not necessarily convex, the theorem may not be
true. 

Additionally, the well-known Wald theorem, namely that all initially
expanding Bianchi models, except type IX, approach the de Sitter space-time 
if all energy conditions are satisfied \cite{Wald:1983ky}, cannot be applied
in the present work, since the scalar field can violate the strong energy
condition. However, there are extensions of the Wald theorem for sources
violating the strong energy condition, known as ``Cosmic No-Hair theorem''
(see \cite{Kitada:1992uh,Kitada:1992bn,Kitada:1991ih}), where in particular
a scalar field with an exponential potential is introduced in addition to a
matter field satisfying the strong and dominant energy conditions,  and all
initially expanding Bianchi models except type IX, are proved to result to
isotropic power-law or exponential inflationary solutions depending on the
potential's slope values (the shear, the 3-curvature and all
components of the energy-momentum tensor of the matter field are strongly
suppressed by the scalar field potential). Moreover, the ``Cosmic No-Hair
theorem'' can be extended in non-minimal scalar field too
\cite{Capozziello:1994nj,Capozziello:1996vp}. Nevertheless, the ``Cosmic
No-Hair
theorem'' has been proven only for constant or exponential potential.

Finally, note that in the case of an additional coupling of the scalar to a
vector field, inflation with an anisotropic stable hair is cosmologically
viable  for the  case of Bianchi I
\cite{Watanabe:2009ct,Kanno:2009ei,Kanno:2010nr,
Watanabe:2010fh,Emami:2010rm,Dimopoulos:2010xq,Dimopoulos:2011ym,Do:2011zza,
Emami:2011yi,Watanabe:2010bu}, as well as for LRS Bianchi II, III and
Kantowski-Sachs geometries   \cite{Hervik:2011xm}. However, these studies are
also restricted to exponential or power-law potentials.

One can go beyond the above specific investigations, applying the powerful
method of dynamical analysis and extracting the asymptotic behavior, as we do
in the present work. In this way one can explicitly see whether the late-time
stable points are always isotropic or not. In particular, as we showed, for
expanding LRS Bianchi I models, the late-time attractors are always
isotropic. Note that
the fact that we may have stable attractors that are not always de Sitter
(solutions dominated by the scalar field, or scaling solutions) is
consistent with the non-validity of the Wald's theorem in our case.
The same results hold for the Kantowski-Sachs models which satisfy ${}^{(3)}\!
R>0$, and in particular we found  that all stable
expanding solutions are isotropic.
However, in the case of expanding LRS Bianchi III models there exist two
anisotropic stable late-time attractors, in addition to the
isotropic ones that exist in Bianchi I too. In particular, we found
the
stable fixed points $R_7(s^*)$ and  $R_{10}(s^*)$, which maintain a non-zero
anisotropy even at asymptotically late times. However, these points, apart
from being inconsistent with observations ($R_7(s^*)$ for usual
dust dark matter has $w_{DE}=0$, while  $R_{10}(s^*)$ has $\Omega_m=0$),
they are always non-accelerating. \footnote{Note   that, as we show in Table \ref{critBIII}, in order for 
$R_7(s^*)$   or $R_{10}(s^*)$  to be stable $s^*$ and $f'(s^*)$ must have
the same sign, which is
not the case for simple potentials such are the exponential and the power-law
ones, that is why this anisotropized point was not found to be stable in
exponential potentials
\cite{Kitada:1992uh,Kitada:1992bn,Kitada:1991ih}, which is just the ``Cosmic
No-Hair theorem''. One should go to more
complicated potentials, like the inverse hyperbolic sine one:   $V(\phi)=V_0
\sinh^{-1}(\alpha\phi), \alpha>0$
\cite{Ratra:1987rm,Wetterich:1987fm,Copeland:2009be,Leyva:2009zz,
Pavluchenko:2003ge,Sahni:1999gb, UrenaLopez:2000aj}, which leads to
$f(s)=s^2-\alpha^2$, in which case $R_7(s^*)$ is stable. That is, we do
verify the ``Cosmic No-Hair theorem'', which has been proven for exponential
potentials, but we show that its extension to a wide range of potentials is
not possible.}
Therefore, we conclude that in the examined geometries, all the
expanding, accelerating, late-time attractors are always isotropic.

Summarizing, applying the phase-space analysis it is possible to prove that
for
LRS Bianchi I, III and Kantowski-Sachs models, the late-time accelerating
attractors are isotropic, at least in the finite region of the phase space. 
A complete investigation of these features should
include the examination of the behavior at infinity since the variable $s$
can diverge { (we recall that the other normalized variables are bounded)}. This
can be
done using the Poincar\'e projection method as in  
\cite{Byland:1998gx,Xu:2012jf}, however for simplicity we do not proceed to
such a detailed analysis in the present work.

\subsubsection{Bounce behavior} 

Let us now investigate another issue of great physical importance, namely
whether anisotropic scalar field cosmology, in the framework of General
Relativity, allows for the bounce realization. Note that the bounce is
impossible in flat FRW scalar-field cosmology, in the case where the scalar
field
is minimally-coupled to General Relativity, since its realization requires
the violation of the null energy condition $\rho+p\geq0$, which cannot be
obtained by a
canonical scalar-field \cite{Novello:2008ra,Cai:2011bs,Qiu:2013eoa}.

Following the approach and conventions of \cite{Solomons:2001ef}, we 
first define the bounce realization. Introducing the scale factors
$X=\ex(t)^{-1}$ and $Y=\ey(t)^{-1}$, and  the corresponding 
expansion parameters $ 
H_x=\dot X/X$ and $ H_y=\dot Y/Y$, a bounce in $X$ occurs at time $t=t_0$ if
and only if $H_x(t_0)=0$ and $\dot H_x
(t_0)>0$, while a bounce in $Y$ occurs at time $t=t_1$ if and only if
$H_y(t_1)=0$ and
$\dot H_y (t_1)>0$. Although in principle it may be possible to have a
bounce in one of the scale factors but not in the other, since this does not
lead to a new expanding region in the universe, in the following we assume
hat the bounce occurs in both $X$ and $Y$ scale factor, but not necessarily
at the same time  \cite{Solomons:2001ef}. Finally, we assume that matter
obeys the null energy condition  
$\rho_m+ p_m\geq 0$.

Defining the 3-curvature scalar as ${}^{(3)}\! R\equiv \frac{2
k}{Y^2}=2 k K,$ where $k=+1, 0, -1$, for Kantowski-Sachs, LRS Bianchi I and
LRS Bianchi III respectively, it is easy to re-write the Raychaudhuri and
the Friedmann equations, and
the evolution equation for anisotropies, in a unified way as: 
\begin{align}
&\dot H=-H^2-2 \sigma_+^2-\frac{1}{6}\left(\rho_m+3
p_m\right)-\frac{1}{3}\dot\phi^2+\frac{1}{3}V(\phi),\nonumber\\
&{}^{(3)}\! R=\dot\phi^2+2 V(\phi)-6 H^2 +6\sigma_+^2+
2\rho_m,\nonumber\\
&\dot\sigma_+=-3 H\sigma_+ -\frac{1}{6}{}^{(3)}\! R \label{BOUNCE}.
\end{align}
Using that $\sigma_+=-\frac{1}{3}\left(H_x-H_y\right)$ and $
H=\frac{1}{3}(H_x+2 H_y)$, as well as the matter equation of state
$p_m=(\gamma-1)\rho_m$, equations (\ref{BOUNCE}) lead to
\begin{equation}\label{BOUNCE2}
\dot\phi^2=-\gamma\rho_m-2\left(\dot H_y+H_y^2-H_x H_y\right).
\end{equation}

Let us assume that we have a bounce in the $Y$ direction at the time
$t=t_1$ (that is   $H_y(t_1)=0$ and $\dot H_y(t_1)>0$). In this case 
\eqref{BOUNCE2} gives 
\begin{equation}\label{BOUNCE3}
\dot\phi^2(t_1)=-\gamma\rho_m(t_1)-2\dot H_y(t_1).
\end{equation} 
{{Thus, taking into account the matter's null energy
condition (that is
$\gamma\rho_m\geq0$) }} we deduce that at the bounce  $\dot\phi^2(t_1) <0$,
that is the 
bounce in the $Y$ direction, and thus the total bounce, is impossible for
real scalar fields.

Therefore, we conclude that {\it{in LRS Bianchi type I, III
and Kantowski-Sachs geometries, with a minimally coupled scalar field to
General
Relativity and a perfect fluid satisfying the null 
energy condition, a cosmological bounce is not permitted (unless the reality
condition is violated)}}. This result is an extension of the no-bounce
theorem proved in   reference \cite{Solomons:2001ef}, if we include matter
additionally to the scalar field. We stress that this result holds only in
General Relativity, since going to modified gravity constructions a bounce
is possible in both isotropic and  anisotropic geometries
\cite{Novello:2008ra,Cai:2011bs,Saridakis:2007cf,Cai:2010zma,Cai:2011tc}.

\section{Application 1: Exponential potential with a cosmological
constant}\label{exppotential}

In the previous section we performed a complete dynamical analysis of
anisotropic scalar-field cosmology, using the advanced f-devisers method,
which allowed us to extract the results without specifying the potential
form. In order to give a specific application, in this section we just
substitute the exponential potential with a cosmological constant in the
general results, obtaining the corresponding dynamical behavior without
repeating the whole analysis from the beginning.

We are interested in a potential of the form 
\be
V(\phi)=V_{0}e^{-\lambda\phi}+\Lambda\label{exppot},
\ee 
since exponential potentials have been extensively studied in cosmological
frameworks  
\cite{Burd:1988ss,Aguirregabiria:1993pm,Ibanez:1995zs,
Aguirregabiria:1996uh,Copeland:1997et,Chen:2008ft,Copeland:2009be,
Coley:2000zw,Coley:2000yc,Pavluchenko:2003ge,Barreiro:1999zs,Leon2009,
Aguirregabiria:1993pk,Leon2010,
Halliwell:1986ja,
Ferreira:1997au,Coley:1997nk,
Heard:2002dr,Rubano:2003et,Leon:2012mt}.

As we mentioned in Table
\ref{fsform}, the corresponding $f(s)$ function is given by
\be
f(s)=-s(s-\lambda),
\label{2.11}
\ee
and thus its roots $f(s^{*}) =0$ and the corresponding derivatives
$f'(s^{*})$ read simply
\begin{eqnarray}
\label{expfs1}
&&s^* =0, \quad f'(s^{*}) =\lambda
\label{2.12a}\\
\label{expfs2}
&&s^* =\lambda, \quad f'(s^{*})=-\lambda.
\label{2.12b}
\end{eqnarray}
In summary, all we have to do is to substitute these values in the general
results of section \ref{Dynamicalanalysis}.  Those
expressions that were independent from $s^*$, will be the same in the present
specific potential case too. Thus, only the results that were depending on
$s^*$, $f(s^{*})$ and $f(0)$ will be now specified. Finally, note that since
the potential at hand has two roots $s^*$, the points of the general
analysis depending on $s^*$, will split to two. In the following subsections
we examine the three anisotropic geometries separately.

\subsection{Kantowski-Sachs metric}

In Kantowski-Sachs geometry the results were summarized in Tables
\ref{critKS} and \ref{critKS2} of subsection \ref{KSphase_space}. Inserting
the specific values (\ref{expfs1}),(\ref{expfs2}) we obtain the following.
\begin{figure}[ht]
\begin{center}
\subfigure[]{
\includegraphics[width=7.5cm, height=6.5cm,angle=0]{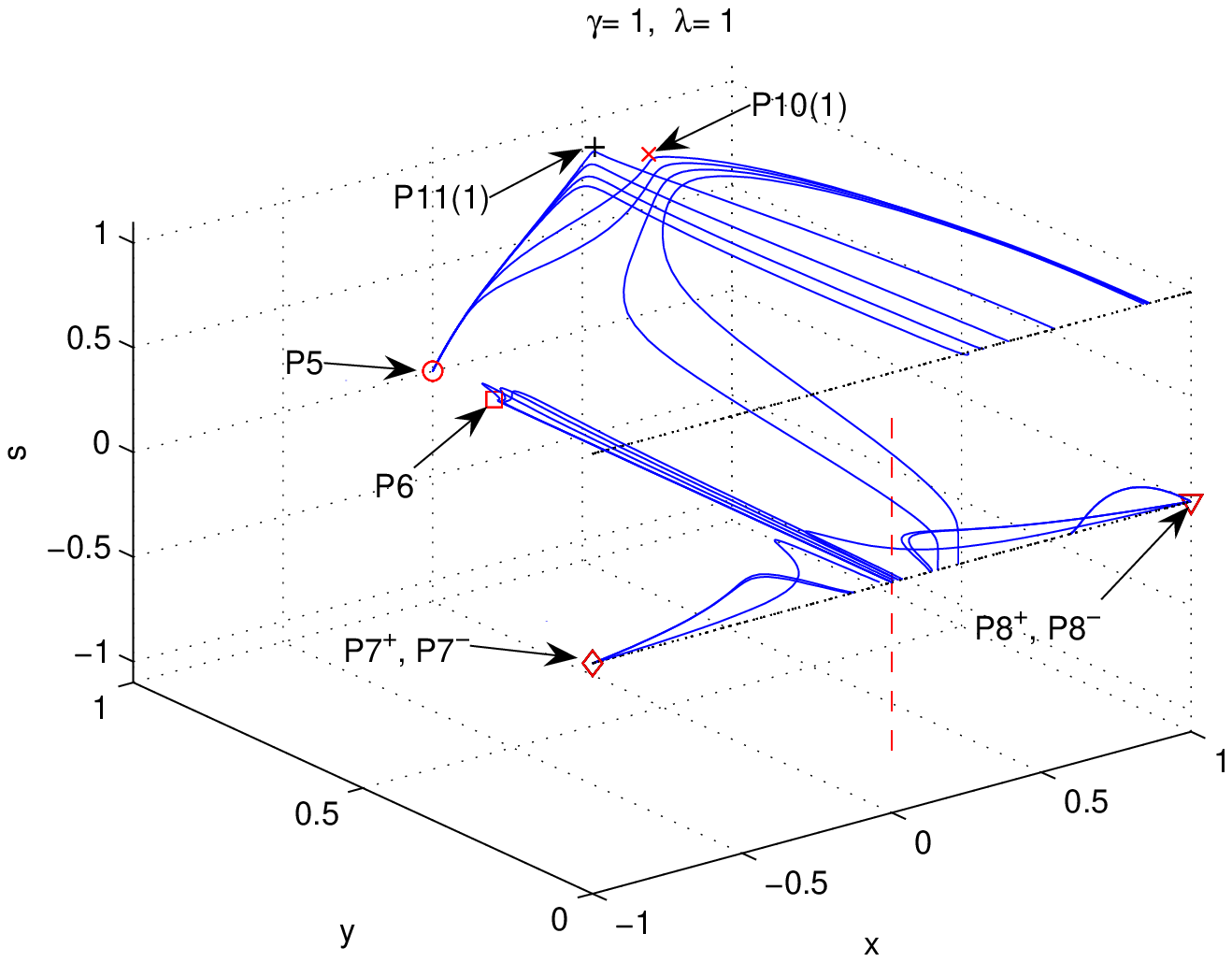}}
\subfigure[]{
\includegraphics[width=7cm, height=6cm,angle=0]{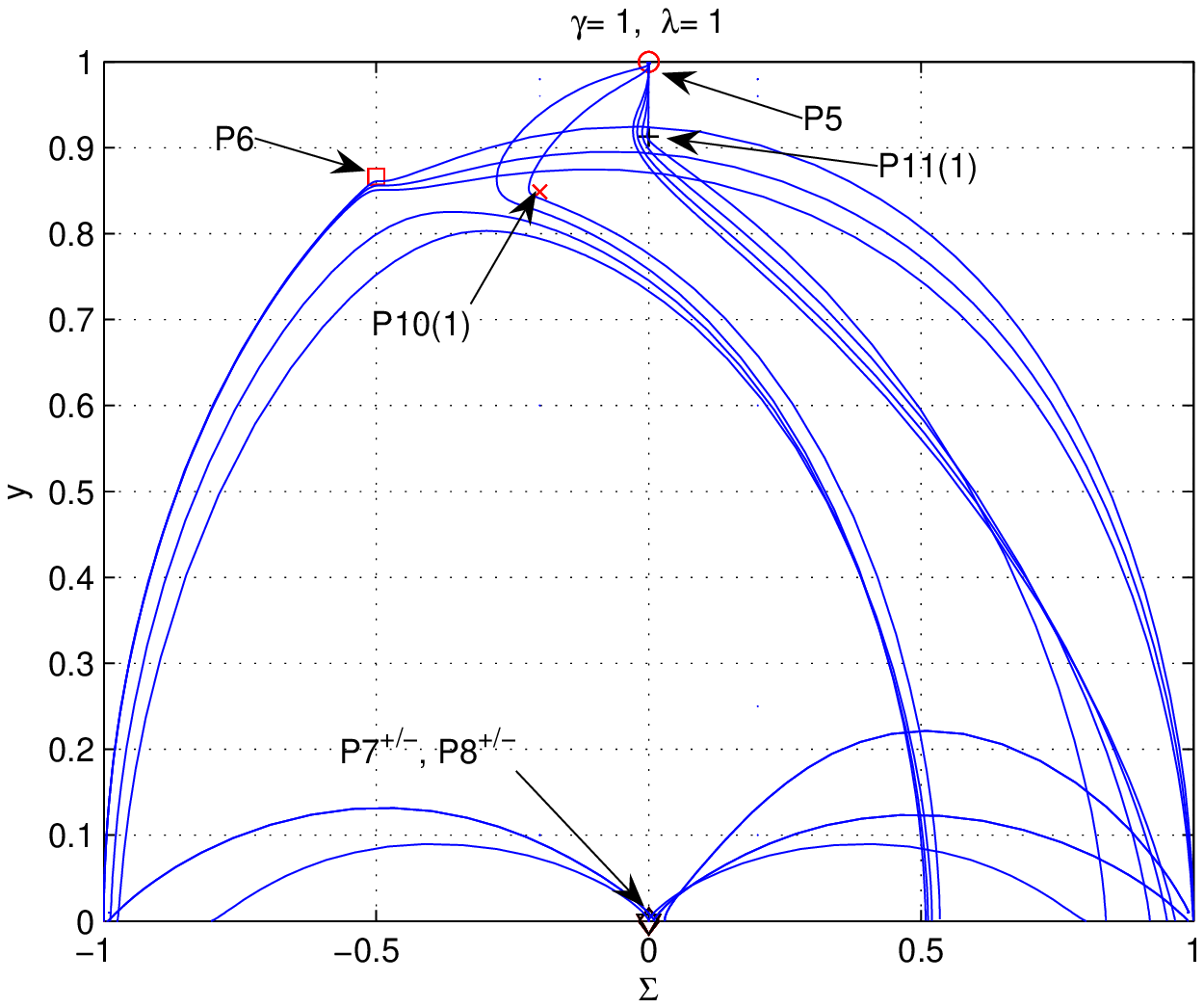}}
\caption{{\label{fig:KSexp}} Projection of the phase space on the:
(a) $x-y-s$ subspace and
(b) $y-\Sigma$ plane, in the case of Kantowski-Sachs geometry with
exponential
potential plus a cosmological constant,   for $\lambda=1$  in units where
$8\pi G=1$, in the case of dust matter
($\gamma=1$). For these parameter choices, and in agreement with
Table  \ref{critKS} and the discussion in the text, depending on the
initial conditions (which determine the basin of attraction) the universe at
late times can result either in the isotropized, dark-energy
dominated, de Sitter point $P_5^+$, or in the contracting solution  $P_8^-$.
Note that  
since in the projection of the graphs, the axis $Q$ that distinguishes
expanding ($Q>0$) from contracting ($Q<0$) solution is absent, there is a 
degeneracy and all expanding points coincide with their contracting
partners.}
\end{center}
\end{figure}

Concerning the physically interesting critical point  $P_5^+$, we see that
it now becomes non-hyperbolic, since one of its eigenvalues becomes zero
since $f(0)=0$ (see \ref{Eigenvalues1}). In order to analyze its
stability we apply the center manifold analysis \cite{wiggins} and we
introduce the coordinate transformation 
\begin{equation}\label{centerP5exp}
 u=\frac{\sqrt{6}}{6}s, v_1=x-\frac{\sqrt{6}}{6}s, v_2=\Sigma-2(Q-1), 
v_3=y-1, v_4=Q-1,
\end{equation} 
finding that the center manifold is given by the graph
\begin{align}\label{centerP5exp2}
&\{(u,v_1,v_2,v_3,v_4): v_1=\sqrt{\frac{2}{3}} \lambda  u^2+2 \left(\lambda
^2-1\right) u^3+{\cal O}(u)^4, v_2={\cal O}(u)^4,\nonumber \\   & \qquad 
\qquad \qquad v_3=-\frac{u^2}{2}-\sqrt{\frac{2}{3}} \lambda 
   u^3+{\cal O}(u)^4, v_4={\cal O}(u)^4, |u|<\delta\},
\end{align}
where $\delta$ is a sufficiently small positive constant. 
The dynamics on the center manifold obeys the gradient-like equation
$
u'=-\sqrt{6} \lambda  u^2+\left(6-2 \lambda ^2\right) u^3+{\cal O}(u)^4$,
and thus the center manifold of the origin is locally asymptotically
of saddle type,  with an 1D
center manifold and 4D stable manifold  \footnote{For $\lambda>0$
($\lambda<0$)  the orbits with    
initially $u>0$ ($u<0$) tend asymptotically to the origin as time passes,
while orbits with
initially $u<0$ ($u>0$) depart from the origin and become unbounded.}.
Interestingly enough, although the 
point $P_5^+$ is in general stable for general potentials (except in the
special case $f(0)=0$), it is not stable
anymore in the specific case of exponential potential with a cosmological
constant  (where we have exactly $f(0)=0$), although it has a
high-dimensional stable manifold and thus, for $\lambda>0$ ($\lambda<0$) it
can
still attract the majority of the orbits having $s>0$ ($s<0$).

The physically interesting critical point  $P_{11}^+(s^*)$  and $P_{12}^+(s^*)$ do not
exist anymore for the first solution  $s^*=0$ of (\ref{expfs1}). For the
second solution $s^*=\lambda$ of (\ref{expfs2}), they become saddle points
(since the stability conditions of Table \ref{critKS} cannot be fulfilled),
and thus they cannot be the late-time state of the universe. Note that in
the case where the cosmological constant is zero, the points  $P_{11}^+(s^*)$  and
$P_{12}^+(s^*)$ become conditionally stable again, and they are the usual
isotropized, quintessence-like and stiff dark-energy points of the
literature \cite{Copeland:1997et}. However, a non-zero
cosmological constant  always dominates over the scalar-field terms at
late times, that is $P_{11}^+(s^*)$ and $P_{12}^+(s^*)$ become saddle, and the
possible expanding attractor is the cosmological-constant, de Sitter solution
 $P_5^+$ \footnote{Note that  $P_5^+$ for $\lambda>0$ ($\lambda<0$)   is a
local attractor for $s>0$ ($s<0$), while
for $s<0$ ($s>0$) the orbits can be unbounded along the $s$-direction at late
times.}.

Finally, as we mentioned in the general-potential analysis of
Kantowski-Sachs geometry of subsection \ref{KSphase_space}, the
contracting points $P_7^-$ or $P_8^-$ become stable for particular
parameter values.

Now, in order to present the aforementioned results in a more
transparent way, we perform a numerical elaboration of our
cosmological system. In Fig. \ref{fig:KSexp} we depict the
projection of the phase space on the $x-y-s$ subspace and  on the $y-\Sigma$
plane, in the case of dust
matter ($\gamma=1$). For these parameter choices, and in agreement with
Table  \ref{critKS} and the discussion in the text, depending on the
initial conditions, which determine the basin of attraction, the universe at
late times can result either in the isotropized, dark-energy
dominated, de Sitter point $P_5^+$, or in the contracting solution  $P_8^-$.

As we see, through the application of our general analysis in this specific
potential, we re-obtained the results of the literature
\cite{Byland:1998gx}, without the need of any
calculation, which reveals the capabilities of the our method.

We close this subsection by stressing that, as we see,
some of the stable points that exist for
general potentials, are not stable or do not even exist in the specific case
of exponential potential with a cosmological constant. This feature may be an
indication that the exponential potentials restrict the dynamics in
scalar-field cosmology.

\subsection{LRS Bianchi III and Bianchi I metrics}

In LRS Bianchi III and Bianchi I geometries the results for a general potential were
summarized
in Tables \ref{critBIII},  \ref{critBIIIb} and \ref{critBIII2} of subsection
\ref{BIIIphase_space}. Inserting the specific values
(\ref{expfs1}),(\ref{expfs2}) of the exponential with cosmological constant
we obtain the following.

The physically interesting critical point  $R_5$  now becomes
non-hyperbolic, since one of its eigenvalues becomes zero
since $f(0)=0$ (see \ref{Eigenvalues3}). 
In order to analyze its
stability we apply the center manifold analysis \cite{wiggins} and {we
introduce the coordinate transformation
\begin{eqnarray}\label{centerR5exp}
&u=\frac{\sqrt{6}}{6}s, v_1=x-\frac{\sqrt{6}}{6}s, v_2=-x^2-(y-1)^2-\Omega_k, \nonumber\\
& v_3=[1-x^2-y^2-\Sigma-\Omega_k], \nonumber\\
& v_4=y-1+\frac{[1-x^2-y^2-\Sigma-\Omega_k]}{2},
\end{eqnarray} 
}
finding that the center manifold is given by the graph
\begin{align}\label{centerR5exp2}
&\{(u,v_1,v_2,v_3,v_4): v_1=\sqrt{\frac{2}{3}} \lambda  u^2+2 \left(\lambda
^2-1\right) u^3+{\cal O}(u)^4,  
  v_2=-u^2-\frac{2\sqrt{6}}{3}u^3+{\cal O}(u)^4,\nonumber \\ & \qquad  \qquad
\qquad
 v_3={\cal O}(u)^4,  v_4=-\frac{u^2}{2}-\sqrt{\frac{2}{3}} \lambda 
   u^3+{\cal O}(u)^4, |u|<\delta\},
\end{align}
where $\delta$ is a sufficiently small positive constant. 
The dynamics on the center manifold obeys the gradient-like equation
$u'=-\sqrt{6} \lambda u^2+\left(6-2 \lambda ^2\right) u^3+{\cal O}(u)^4$,
and thus the center manifold of the origin is locally asymptotically
of saddle type, with an 1D center manifold and 4D stable manifold. 
However, since  it has a high-dimensional stable manifold,  for $\lambda>0$
($\lambda<0$) it can still
attract the majority of the orbits having $s>0$ ($s<0$) initially. { Note
that for $R_5$ the center manifold analysis and the stability conditions
are the same for the LRS Bianchi III and Bianchi I cases.}
 \begin{figure}[ht]
\begin{center}
\includegraphics[width=9cm, height=8cm,angle=0]{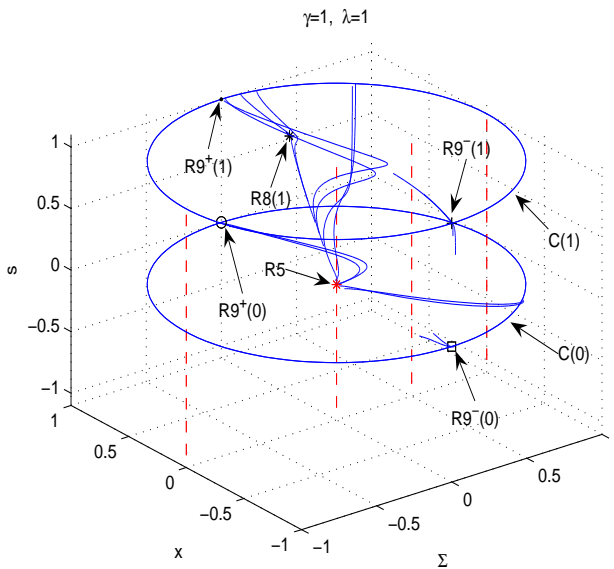}
\caption{{\label{fig:BIIIexp}} Orbits in the phase space, in the case of
 LRS Bianchi III geometry with exponential potential plus a cosmological
constant, for $\lambda=1$ in units where $8\pi G=1$, in the case of dust
matter
($\gamma=1$).  At late times the universe results in the isotropized, 
dark-energy dominated, de Sitter solution $R_5$. For completeness, we
also depict the circles corresponding to the two curves of critical
points $C(0)$ and $C(1)$, that is for $s^*=0$ and $s^*=\lambda=1$,
respectively (see Table \ref{critBIII}).}
\end{center}
\end{figure}

  The physically interesting critical points  $R_6(s^*)$, $R_7(s^*)$, $R_{8}(s^*)$ and $R_{10}(s^*)$ do not
exist for the first solution  $s^*=0$ of (\ref{expfs1}). For the second
solution $s^*=\lambda$ they all become saddle points since $s^* f'(s^*)=-\lambda^2<0$, and thus they cannot
be the late-time state of the universe (unless the cosmological
constant is zero, in which case they remain stable).

As we see, through the application of our general analysis in this specific
potential, we re-obtained the results of the literature
\cite{Byland:1998gx}, without the need of any
calculation, which reveals the capabilities of   our method.

Similarly to the Kantowski-Sachs case, we mention that
some of
the stable
points that exist for general potentials, are not stable in the specific case
of exponential potential with a cosmological constant. This feature may be an
indication that the exponential potentials restrict
the dynamics in scalar-field cosmology.

 Now, in order to present the aforementioned results in a more transparent
way, we perform a numerical elaboration of our cosmological system. In Fig.
\ref{fig:BIIIexp} we depict some orbits  in the phase space, in the case of
dust matter ($\gamma=1$).

\begin{figure}[ht]
\begin{center}
\includegraphics[width=9cm, height=8cm,angle=0]{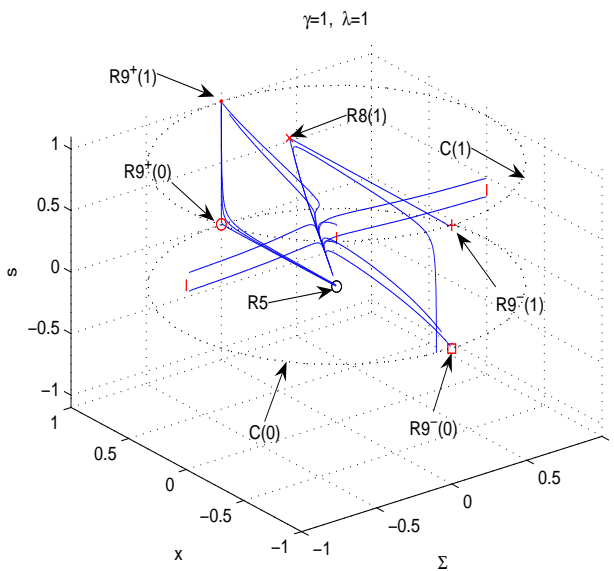}
\caption{{\label{fig:BIexp}} Orbits in the phase space, in the case of
LRS Bianchi I geometry with exponential potential plus a cosmological
constant, 
 for  $\lambda=1$  in units where $8\pi G=1$, in the case of dust matter
($\gamma=1$). At late times the universe results in the isotropized, 
dark-energy dominated, de Sitter solution $R_5$. For completeness, we
also depict the dotted circles corresponding to the two curves of critical
points $C(0)$ and $C(1)$, that is for $s^*=0$ and $s^*=\lambda=1$,
respectively (see Table \ref{critBIII}).}
\end{center}
\end{figure}

Additionally, in Fig.
\ref{fig:BIexp} we depict some orbits  in the phase space, in the case of
dust matter ($\gamma=1$) for the boundary set of LRS Bianchi I geometry.

\section{Application 2: Power-law potential}
\label{powerlawpot}

In this section we give a second specific application, of the
general-potential analysis of section \ref{Dynamicalanalysis}. In
particular, we are interested in the power-law potential of the form
\cite{Ratra:1987rm,Abramo:2003cp,Aguirregabiria:2004xd,
Copeland:2004hq,Saridakis:2009pj,
Saridakis:2009ej,Leon:2009dt,Chang:2013cba,Skugoreva:2013ooa,Pavlov:2013nra,
Peebles:1987ek}
\begin{equation}
\label{powerpot}
V(\phi)=V_0 \phi^n.
\end{equation}
As far as we are aware,  the potential 
\eqref{powerpot} in the anisotropic context has not been investigated up to
now. 
 
In this particular case, the use of the general definitions
(\ref{sdef}),(\ref{fdef}) gives $s=-\frac{n}{\phi}$ and
$f=-\frac{n}{\phi^2}$ and thus
\be
f(s)=-\frac{s^2}{n},
\label{2.12}
\ee
and therefore its roots $f(s^{*}) =0$ and the corresponding derivatives
$f'(s^{*})$ read simply
\begin{eqnarray}
\label{powerfs1}
&&s^* =0, \quad f'(s^{*}) =0.
\end{eqnarray}
In summary, all we have to do it to substitute these values in the general
results of section \ref{Dynamicalanalysis}.  Those
expressions that were independent from $s^*$, will be the same in the present
specific potential case too. Thus, only the results that were depending on
$s^*$, $f(s^{*})$ and $f(0)$ will be now specified.   In the following
subsections
we examine the three anisotropic geometries separately.

\subsection{Kantowski-Sachs metric}

In Kantowski-Sachs geometry the results were summarized in Tables
\ref{critKS} and \ref{critKS2} of subsection \ref{KSphase_space}. Inserting
the specific values (\ref{powerfs1})  we obtain the following.

Concerning the physically interesting critical point  $P_5^+$, we see that
it now becomes non-hyperbolic, since one of its eigenvalues becomes zero
since $f(0)=0$ (see \ref{Eigenvalues1}). In order to analyze its
stability we apply the center manifold analysis \cite{wiggins} and we
introduce the coordinate transformation 
 \eqref{centerP5exp},
finding that the center manifold is given by the graph
\begin{equation}
\{(u,v_1,v_2,v_3): v_1=-\frac{2 u^3}{n}  +{\cal O}(u)^4, v_2={\cal O}(u)^4, 
v_3=-\frac{u^2}{2}+{\cal O}(u)^4, v_4={\cal O}(u)^4, |u|<\delta\},
\end{equation}
where $\delta$ is a sufficiently small positive constant. 
The dynamics on the center manifold obeys the gradient-like equation
$
u'=\frac{6}{n} u^3+{\cal O}(u)^5$,
and thus  for $n>0$  the center manifold of the origin is locally
asymptotically of saddle type,  with  1D
center manifold and 4D stable manifold, while for $n<0$ point $P_5^+$ is
completely
stable. Furthermore, the other two physically interesting critical points,
namely $P_{11}$  and $P_{12}$, do not
exist any more, since $s^*=0$.  
\begin{figure}[ht]
\begin{center}
\includegraphics[width=9cm, height=8cm,angle=0]{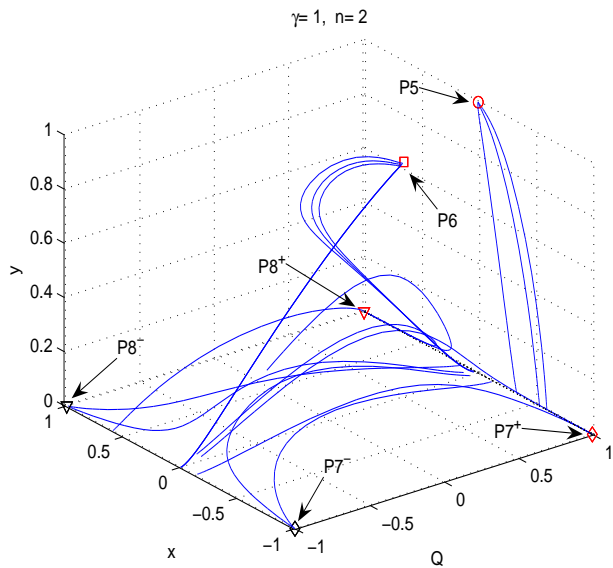}
\caption{{\label{fig:KSn4}} Orbits in the phase space, in the case of
Kantowski-Sachs geometry with power-law potential, for $n=2$, in the case of
dust matter ($\gamma=1$). For these parameter choices, and in agreement with
Table  \ref{critKS} and the discussion in the text, depending on the
initial conditions (which determine the basin of attraction) the universe at
late times can result either in the isotropized, dark-energy
dominated, de Sitter point $P_5^+$, or in the contracting solutions  $P_7^-$
or $P_8^-$.}
\end{center}
\end{figure}

Finally, the contracting points $P_7^-$ and $P_8^-$ become non-hyperbolic
too. Applying the   center manifold analysis  we introduce the coordinate
transformation 
\begin{equation}\label{centerP78exp}
 u_1=\frac{1}{2}(2\Sigma-Q-1),u_2=s, v_1=Q+1,  v_2=x-\eta, v_3=y,
\end{equation} where $\eta=-1$ for $P_7^-$ and  $\eta=+1$ for $P_8^-$,
finding that   the center
manifold is given by the graph
\begin{equation}
\{(u_1,u_2,v_1,v_2,v_3): v_1=O(u_1^3),  v_2=-\tfrac{\eta}{2} u_1^2+O(u_1)^3, 
v_3=O(u_1)^3, |(u_1,u_2)|<\delta\}.
\end{equation} 
  The dynamics on the
center manifold of $P_{7, 8}^-$ is dictated  by
$
u_1'=O(u_1^4), u_2'=\tfrac{\sqrt{6}\eta}{n} u_2^2+O(u_1^2 u_2^2).
$  
Hence, for $\eta=-1$, that is for $P_7^-$, and for 
$n>0$ ($n<0$) the orbits with    
initially $u_2>0$ ($u_2<0$) tend asymptotically to the origin as   time
passes, while orbits with
initially $u_2<0$ ($u_2>0$) depart from the origin and become unbounded along
the $u_2$-axis.
For $\eta=+1$, that is for $P_8^-$, and for 
$n<0$ ($n>0$) the orbits with    
initially $u_2>0$ ($u_2<0$) tend asymptotically to the origin as   time
passes, while orbits with
initially $u_2<0$ ($u_2>0$) depart from the origin and become unbounded along
the $u_2$-axis. In summary $P_7^-$ and $P_8^-$ possess a 3D stable
manifold and a 2D center manifold. Thus, although in the general case $P_7^-$
and $P_8^-$ can be stable (with the exception of $f'(0)=0$),  they are not
completely stable in this special case of the power-law potential (where we
have exactly $f'(0)=0$). However, they can still
conditionally attract  the majority of the orbits.

In order to present the above results in a more transparent way, we
perform a numerical elaboration of our cosmological system. In Fig.
\ref{fig:KSn4} we depict some orbits  in the phase space, in the case of
dust matter ($\gamma=1$). For these parameter choices, and in agreement with
Table  \ref{critKS} and the discussion in the text, depending on the
initial conditions, which determine the basin of attraction, the universe at
late times can result either in the isotropized, dark-energy dominated, de
Sitter point $P_5^+$, or in the contracting solutions  $P_7^-$ or $P_8^-$.

Finally, and similarly to the previous section, we see here too that some of
the stable points that exist for general potentials, do not exist in the
specific case of power-law potential. This feature may be an indication that
  power-law potentials   restrict the dynamics in scalar-field
cosmology.

\subsection{LRS Bianchi III and Bianchi I metrics}

In LRS Bianchi III and Bianchi I geometries the results for a general potential were
summarized
in Tables \ref{critBIII} and \ref{critBIII2} of subsection
\ref{BIIIphase_space}.  
Inserting the specific values (\ref{powerfs1}) of the
power-law potential we obtain the following.

The physically interesting critical point  $R_5$  now becomes
non-hyperbolic, since one of its eigenvalues becomes zero
since $f(0)=0$ (see \ref{Eigenvalues3}). 
In order to analyze its
stability we apply the center manifold analysis \cite{wiggins} and we
introduce the coordinate transformation
\eqref{centerR5exp},
finding that the center manifold is given by the graph
\begin{align}\label{centerR5exp22}
& \{(u,v_1,v_2,v_3,v_4): v_1= -\frac{2 u^3}{n}+{\cal O}(u)^4,   v_2= -u^2
+{\cal O}(u)^4,
 v_3={\cal O}(u)^4, \nonumber \\ & \qquad  \qquad \qquad 
v_4=-\frac{u^2}{2}+{\cal O}(u)^4, |u|<\delta\},
\end{align}
where $\delta$ is a sufficiently small positive constant. 
The dynamics on the center manifold obeys the gradient-like equation
$
u'=\frac{6}{n} u^3+{\cal O}(u)^5$,
and thus  for $n>0$  the center manifold of the origin is locally
asymptotically of saddle type,  with an 1D center manifold and 4D stable
manifold, while for $n<0$ point $R_5$ is completely stable.  
However, since  it has a high-dimensional stable manifold  it can still
attract the majority of the orbits.  {Note
that for $R_5$ the center manifold analysis and the stability conditions
are the same for the LRS Bianchi III and Bianchi I cases.}
\begin{figure}[ht]
\begin{center}
\includegraphics[width=9cm, height=8cm,angle=0]{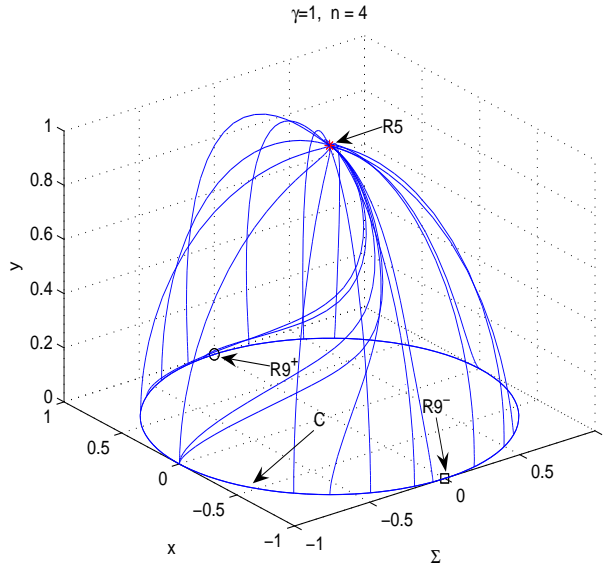}
\caption{{\label{fig:BIIIn2}} Orbits in the phase space, in the case of
LRS Bianchi III geometry with power-law potential, for $n=4$, in the case of
dust
matter ($\gamma=1$). At late times the universe results in the isotropized,
dark-energy dominated, de Sitter point
$R_5$. For completeness, we
also depict the  circle corresponding to the  curves of critical
points $C$ (see Table \ref{critBIII}).}
\end{center}
\end{figure}
 
 Finally, the other three physically interesting critical points,
namely $R_6(s^*)$, $R_7(s^*)$, $R_8(s^*)$, and $R_{10}(s^*)$ do not exist any more, since $s^*=0$.
Similarly to the previous   Kantowski-Sachs case, we see
that some of the stable points that
exist for general potentials, do not exist in the specific case of power-law
potential. This feature may be an indication that    power-law potentials
  restrict the dynamics in scalar-field cosmology.

 Lastly, we perform a numerical elaboration of our cosmological system and in
Fig. \ref{fig:BIIIn2} we depict some orbits  in the phase space, in the case
of dust matter ($\gamma=1$).

\section{Conclusions}
\label{conclusions}

In this work we performed a detailed dynamical analysis of anisotropic
scalar-field cosmologies, and in particular of the most significant
Kantowski-Sachs, LRS Bianchi I and LRS Bianchi III cases. These geometries
are
very interesting, especially if one finds late-time isotropized solutions,
since they provide an explanation for the observed universe isotropy,
instead of having to assume it from the start, like in FRW cosmology.
Additionally, the whole discussion may be relevant for the explanation of
the anisotropic ``anomalies'' reported in the recently announced Planck
probe results   \cite{Ade:2013zuv}.

The investigation of the phase-space behavior is very helpful and necessary
in every cosmology, since it bypasses the non-linearities and complications
of the cosmological equations, which prevent complete analytical treatments,
and it reveals the late-time behavior of the universe. However, a 
disadvantage of the relevant literature up to now was that the whole analysis
had to be performed separately and from the start, for every different
scalar-field potential
\cite{Burd:1988ss,REZA,Copeland:1997et,Coley:2003mj,Leon2011,Gong:2006sp,
Setare:2008sf,
Chen:2008ft,Gupta:2009kk,
Matos:2009hf,
Copeland:2009be,Leyva:2009zz,
Farajollahi:2011ym,UrenaLopez:2011ur,Escobar:2011cz,
Escobar:2012cq,Xu:2012jf,Leon:2013qh}. Apart from the huge calculation needs,
this
feature put limits in the whole approach, weakening the idea of the dynamical
analysis itself, since the results were potential-dependent.

Therefore, in the present work we extended beyond the usual procedure, and we
followed the powerful method of the $f$-devisers, which allows us to perform
the whole analysis without the need of an  {\it{a priori}}  specification of
the potential. In particular, with the introduction of the new variables $f$
and $s$, one adds an extra direction in the phase-space, whose
neighboring points correspond to ``neighboring'' potentials. Thus, after the
general analysis has been completed, the substitution of the specific $f(s)$
for the desired potential gives immediately the specific results, through a
form of intersection of the extended phase-space. That is, one can just
substitute the specific potential form in the final results and obtain the
corresponding behavior, instead of having to repeat the whole dynamical
elaboration from the start. This general investigation appears for the first
time in the literature and it eliminates almost any calculation need from
future relevant investigations.

Following the aforementioned extended procedure, we extracted and we
presented the complete and detailed phase-space behavior of Kantowski-Sachs,
LRS Bianchi I and LRS Bianchi III geometries for a wide range of scalar-field
potentials,
calculating
also the values of basic observables such is the the various density
parameters, the   deceleration parameter and the dark-energy and total
equation-of-state parameters. We found a very rich behavior: in all cases the
universe can result in isotropized solutions with observables in agreement
with observations, independently of the initial conditions and its specific
evolution. In particular, all
expanding, accelerating, stable attractors  are  isotropic.  Amongst
others, the universe can result in a dark-energy
dominated, accelerating, de Sitter solution, in a dark-energy dominated
quintessence-like solution, or in a stiff-dark energy solution with
dark-energy and dark-matter densities of the same order.

Additionally, we proved that in the examined geometries, namely in
LRS Bianchi type I, III
and Kantowski-Sachs models, with a real, minimally-coupled scalar field to
General
Relativity and a perfect fluid satisfying the null 
energy condition, a cosmological bounce is impossible.

Finally, for completeness, after the extraction of the general-potential
results, and in order to show how the analysis is applied to specific
potentials, we presented the application to two well-studied potentials,
namely the exponential plus cosmological constant one and the power-law one.
In the first case we re-obtained the results of the literature, while in the
second case, the relevant analysis appears for the first time. However, the
interesting point is that, as we showed, some of the stable points that
exist for general potentials, are not stable or do not even exist in the two
specific  potentials. This feature may be an indication that the exponential
and power-law potentials might restrict the dynamics in scalar-field
cosmology, opening the way to the introduction of more complicated
ones.

\section*{Acknowledgments}

The authors wish to thank S. Basilakos,  Y.~F.~Cai, A. A. Coley, S. Dutta, G.
Kofinas, T. Koivisto,
D. F. Mota, 
L. Perivolaropoulos, P. Sandin,
R. Sussman and D.
Wands for useful comments and to J. Barrow for drawing our attention to
useful references. Thanks are due to C. Uggla for useful comments and for bring our attention to related works. Additionally, the authors desire to thank   two
unknown referees for valuable comments and suggestions. 
G.L. was partially supported by PUCV through Proyecto DI Postdoctorado 2013,
by COMISI\'ON NACIONAL DE CIENCIAS Y TECNOLOG\'IA through Proyecto FONDECYT
DE POSTDOCTORADO 2014  grant  3140244 and by DI-PUCV grant 123.730/2013.  The
research of
E.N.S. is implemented within the framework of the Operational
Program ``Education and Lifelong Learning'' (Actions Beneficiary:
General Secretariat for Research and Technology), and is co-financed
by the European Social Fund (ESF) and the Greek State.

\begin{appendix}

\section{Eigenvalues of the perturbation matrix $\bf{\Xi}$ for each critical
point}

\subsection{Kantowski-Sachs models}\label{Eigenvalues1}

The system \eqref{KSQ}-\eqref{KSs}
 admits twelve isolated critical points  (six
corresponding to expanding universe and six corresponding  to
contracting one) and  ten
curves of critical points (five corresponding to expanding universe and five
 corresponding  to
contracting one), presented in Table
\ref{critKS}. In this appendix we calculate the eigenvalues
of the perturbation matrix
$\bf{\Xi}$ for each critical point and curve of critical points. We use the
notation
$\epsilon=\pm1$.

For the curves of critical points $P_1^\epsilon$ the
associated eigenvalues read $\left\{3\epsilon, 2\epsilon,
3\epsilon(2-\gamma),0,0\right\}$, while for the curves of critical
points
$P_2^\epsilon$ they are $\left\{6\epsilon, 3\epsilon,
3\epsilon(2-\gamma),0,0\right\}$. That is they are both non-hyperbolic,
behaving as curves of unstable points (for $\epsilon=+1$) or curves of saddle
points  (for $\epsilon=-1$).

For the curves of critical points $P_3^\pm$ the
associated eigenvalues read
\begin{eqnarray*} & \left\{0,\pm \frac{3\gamma}{4-3\gamma}, \mp
\frac{3(2-\gamma)}{4-3\gamma},
\mp\frac{3\left[(2-\gamma)+\sqrt{(2-\gamma)(18-41\gamma+24\gamma^2)}\right]
}{2(4-3\gamma)}, \right. \nonumber \\
& \left. \mp\frac{3\left[(2-\gamma)-\sqrt{(2-\gamma)(18-41\gamma+24\gamma^2)}
\right]}{2(4-3\gamma)}\right\},
\end{eqnarray*}
and thus both curves $P_3^\pm$ are normally hyperbolic (curves of saddle
points) since
the eigenvector associated to the zero eigenvalue is tangent to the
$s$-axis.

For the curves of critical points $P_4^\epsilon$ the associated eigenvalues
are
\begin{eqnarray*} &
\left\{-\frac{3\epsilon}{2}(2-\gamma),-\frac{3\epsilon}{2}(2-\gamma),\frac
{3\gamma\epsilon}{2},-\epsilon(2-3\gamma)\right\},
\end{eqnarray*} 
hence, they
are curves of saddle points.

For the critical points $P_5^\epsilon$ the associated eigenvalues
are
 \begin{eqnarray*}
 &\left\{-3\epsilon, -2\epsilon,
-3\gamma\epsilon, -\frac{3\epsilon}{2}\left[1-\sqrt{1-\frac{4
f(0)}{3}}\right], 
-\frac{3\epsilon}{2}\left[1+\sqrt{1-\frac{4
f(0)}{3}}\right]\right\},
\end{eqnarray*}
and therefore they are asymptotically stable  (for $\epsilon=+1$)
or unstable (for $\epsilon=-1$) provided $f(0)>0.$

For the critical points $P_6^\epsilon$ the associated eigenvalues
are
 \begin{eqnarray*} &
\left\{-3\epsilon, \frac{3\epsilon}{2}, -\frac{3\gamma\epsilon}{2},
-\frac{3\epsilon}{4}\left[1+\sqrt{1-4f(0)}\right],  
-\frac{3\epsilon}{4}\left[1-\sqrt{1-4f(0)}\right]\right\}, 
\end{eqnarray*}
and thus they are saddle points.

For the critical points $P_7^\epsilon$  the associated
eigenvalues write
$$\left\{3\epsilon, 0, 4\epsilon,
3\epsilon(2-\gamma),\sqrt{6}f'(0)\right\},$$ while for   $P_8^\epsilon$
they are $$\left\{3\epsilon, 0, 4\epsilon,
3\epsilon(2-\gamma),-\sqrt{6}f'(0)\right\}.$$ In both cases, as can be seen
by a simple center manifold analysis \cite{wiggins}, the zero
eigenvalue possesses an eigen-direction tangent to the $\Sigma$-axis passing
through the critical point, and the eigenvector associated to the eigenvalue
$\pm\sqrt{6}f'(0)$ is tangent to the $s$-axis. Thus, and assuming
 $\gamma\neq 2$,   for $f'(0)=0$ both $P_7^\epsilon$
and $P_8^\epsilon$ have a 2D center manifold tangent at the corresponding
fixed point to the $\Sigma$-$s$ plane. Moreover, for $f'(0)<0,$ $P_7^+$ 
behaves as saddle point with a   4D stable manifold, while
$P_8^-$ behaves as saddle point with a 4D-unstable manifold. 
Similarly, for $f'(0)>0$ $P_7^-$ behaves as saddle point with a   a 4D
unstable manifold, while
$P_8^-$ behaves as saddle point with a 4D-stable manifold.
Finally, note that the points $P_7^\epsilon$ and
$P_8^\epsilon$ are special points of the curves of critical
points $C_\epsilon(0)$ (see below), and are
given separately for clarity.

For the  critical points $P_9^\epsilon(s^*)$ the associated
eigenvalues write as
\begin{eqnarray*}
&\left\{-\frac{3 \epsilon\left\{(2-\gamma ) s^*+\sqrt{(2-\gamma ) \left[4
\left(3
\gamma ^2-\sqrt{9 \gamma ^4+\Delta_1}\right)+\Delta_2\right]}\right\}}{2 (4-3
\gamma ) s^*},  \right.
\nonumber\\ & \left. 
   -\frac{3\epsilon
\left\{(2-\gamma ) s^*-\sqrt{(2-\gamma )
\left[4
   \left(3 \gamma ^2-\sqrt{9 \gamma
^4+\Delta_1}\right)+\Delta_2\right]}\right\}}
{2 (4-3 \gamma ) s^*}, \right.
\nonumber\\ & \left. 
-\frac{3 \epsilon
   \left\{(2-\gamma ) s^*-\sqrt{(2-\gamma ) \left[4 \left(3 \gamma
^2+\sqrt{9 \gamma ^4+\Delta_1}\right)+\text{$\Delta
   $2}\right]}\right\}}{2 (4-3 \gamma ) s^*}, \right.
\nonumber\\ & \left.  
-\frac{3 \epsilon\left\{(2-\gamma ) s^*+\sqrt{(2-\gamma ) \left[4 \left(3
\gamma
^2+\sqrt{9
   \gamma ^4+\Delta_1}\right)+\Delta_2\right]}\right\}}{2 (4-3 \gamma )
s^*}, \right. \nonumber\\ & \left. -\frac{6 \gamma \epsilon
   f'\left(s^*\right)}{(4-3 \gamma ) s^*}\right\},
\end{eqnarray*}
where $\Delta_1=[\gamma  (3 \gamma -4)+2]^2 \left(s^*\right)^4+2 \gamma ^2
[3 \gamma  (3 \gamma -4)+2]  ({s^*})^2$ and $\Delta_2=[10-(25-12 \gamma )
\gamma] ({s^*})^2$. Hence, they are non-hyperbolic for
$\gamma=\frac{2}{3}$, otherwise they are of saddle
type.

For the critical points $P_{10}^\epsilon(s^*)$ the associated eigenvalues are
\begin{eqnarray*}&\left\{-\frac{3 \epsilon
\left[ ({s^*})^2+2\right]}{ ({s^*})^2+4},-\frac{6 \epsilon  \left[\gamma
    ({s^*})^2- ({s^*})^2+\gamma \right]}{ ({s^*})^2+4},-\frac{6 \epsilon  s^*
   f'\left(s^*\right)}{ ({s^*})^2+4},  \right.
\nonumber\\ & \left. 
	-\frac{3
\epsilon
   \left[ ({s^*})^2+2+\sqrt{\left[2+(s^*)^2\right]\left[18-7 (s^*)^2\right]}\right]}{2
   \left[ ({s^*})^2+4\right]}, \right. \nonumber\\ &\left.
	-\frac{3 \epsilon
\left[ ({s^*})^2+2-\sqrt{\left[2+(s^*)^2\right]\left[18-7 (s^*)^2\right]}\right]}{2 \left[ ({s^*})^2+4\right]}\right\}\nonumber,
   \end{eqnarray*}
{thus they are nonhyperbolic for $({s^*})^2\in \left\{0,2, \frac{\gamma}{1-\gamma}\right\}$ or $f'(s^*)=0.$  For $0<\gamma <2$ and  $-\sqrt{2}<s^*<\sqrt{2}$ they are saddle points.}

For the critical points $P_{11}^\epsilon(s^*)$ the associated eigenvalues
read
\begin{eqnarray*}
&\left\{\frac{1}{2} \epsilon
\left[\left({s^*}\right)^2-6\right],\frac{1}{2} \epsilon
\left[\left({s^*}\right)^2-6\right],\epsilon
   \left[\left({s^*}\right)^2-3 \gamma \right],
	\epsilon
\left[\left({s^*}\right)^2-2\right],-\epsilon  s^*
   f'\left(s^*\right)\right\}.
   \end{eqnarray*}
 Thus, $P_{11}^\epsilon(s^*)$ is non-hyperbolic for ${s^*}^2\in
\{0,2,6, 3\gamma\}$ or $f'(s^*)=0$. Furthermore, $P_{11}^+(s^*)$
($P_{11}^-(s^*)$)  is asymptotically stable (unstable) for $0<\gamma\leq
\frac{2}{3}, -\sqrt{3\gamma}<s^*<0, f'(s^*)<0$ or $0<\gamma\leq
\frac{2}{3}, 0<s^*<\sqrt{3\gamma}, f'(s^*)>0$ or $\frac{2}{3}<\gamma\leq 2,
-\sqrt{2}<s^*<0, f'(s^*)<0$ or $\frac{2}{3}<\gamma\leq 2, 0<s^*<\sqrt{2},
f'(s^*)>0$. In any other case they are saddle points.

For the critical points $P_{12}^\epsilon(s^*)$ the associated eigenvalues are
\begin{eqnarray*}
&\left\{\epsilon(3 \gamma -2),-\frac{3\epsilon}{2} (2-\gamma
),-\frac{3\epsilon}{4} \left\{-\gamma
+\sqrt{(2-\gamma ) \left[\frac{24 \gamma
   ^2}{\left({s^*}\right)^2}-9 \gamma +2\right]}+2\right\},\right.\nonumber\\
& \left.
\frac{3\epsilon}{4} \left\{-\gamma -\sqrt{(2-\gamma ) \left[\frac{24 \gamma
   ^2}{\left({s^*}\right)^2}-9 \gamma
+2\right]}+2\right\},%\right.\nonumber\\ %&\left.
	-\frac{3 \epsilon \gamma  f'\left(s^*\right)}{s^*}\right\}.
\end{eqnarray*}
Therefore, $P_{12}^+(s^*)$ ($P_{12}^-(s^*)$) is asymptotically
stable
(unstable) if
either $0<\gamma<\frac{2}{3},\, s^*<-\sqrt{3\gamma}$ and
$f'(s^*)<0,$ or $0<\gamma<\frac{2}{3},\, s^*>\sqrt{3\gamma}$
and $f'(s^*)>0$, while they are non-hyperbolic if either $\gamma=0,$ or
$\gamma=\frac{2}{3},$ or ${s^*}^2=3\gamma,$ or $s^*=0,$ or
$f'\left(s^*\right)=0$.   In any other case they are saddle points.

Finally, for the curves of critical points $C_\epsilon(s^*)$ the associated
eigenvalues are
\begin{eqnarray*} & \left\{0, 3\epsilon (2-\gamma), -2\left(-2\epsilon+\cos
u\right), 3\epsilon-\sqrt{\frac{3}{2}}s^*\sin u, 
-\sqrt{6}f'(s^*)\sin u\right\}\nonumber
\end{eqnarray*}
and thus $C_\epsilon(s^*)$ are normally hyperbolic and their stability is
determined by the sign of the non zero eigenvalues. Hence, $C_+(s^*)$  are  
unstable for $0\leq \gamma<2,0<u<\pi, s^*<\sqrt{6} \csc (u),
f'\left(s^*\right)<0$ or $0\leq \gamma<2,\pi<u<2\pi, s^*>\sqrt{6} \csc (u),
f'\left(s^*\right)>0$, otherwise they are saddle points. Similarly, 
$C_-(s^*)$ 
are stable for $0\leq \gamma<2, 0<u<\pi, s^*>-\sqrt{6} \csc (u),
f'\left(s^*\right)>0$, or $0\leq \gamma<2,\pi<u< 2\pi, s^*<-\sqrt{6} \csc
(u),
f'\left(s^*\right)<0$,  otherwise they are saddle points.

\subsection{Expanding LRS Bianchi III and Bianchi I models}\label{Eigenvalues3}

The system \eqref{BIIISigma}-\eqref{BIIIs}, in the expanding universe
sub-space, admits six isolated critical points and six curves of critical
points,  presented in Table \ref{critBIII}. We mention that these points and
curves correspond only to the half phase-space of expanding solutions. Thus,
the whole phase-space admits also their symmetric partners corresponding to
contractions, which coordinates and observables are given by the expanding
ones under the transformation (\ref{(78)}). In this appendix we calculate
the eigenvalues of the perturbation matrix $\bf{\Xi}$ for each critical point
and curve of critical points of the expanding solutions.

For the curve of critical  points  $R_1^+$  the associated
eigenvalues read
$\left\{2,3,0,0,3(2-\gamma)\right\}$, while  for   the
curve of critical points  $R_1^-$ they read
$\left\{6,3,0,0,3(2-\gamma)\right\}$. Thus,  they are both
non-hyperbolic, with 3D unstable manifold. { For Bianchi I models the
eigenvalues for $R_1^\pm$ are the same, apart from the first one since the
  extra $\Omega_k$-direction    is not required for the analysis. Thus, the 
stability conditions of Bianchi I are the same as for LRS Bianchi III.}

For the curve of critical points
$R_2$, the associated
eigenvalues are
$\left\{-\frac{3}{2}, -\frac{3}{2},\frac{3}{2},0,
3(1-\gamma)\right\},$
and thus they are
non-hyperbolic (the zero-eigenvalue is associated with an eigenvector
tangent to the $s$-axis), behaving as saddle. { This point does not belong
to the LRS Bianchi I boundary set.}

For the curve of critical points $R_3$ the  
eigenvalues are $\left\{3 \gamma -2,0,\frac{3 \gamma }{2},\frac{3
\gamma }{2}-3,\frac{3 \gamma }{2}-3\right\},$ and thus they are
non-hyperbolic (the zero-eigenvalue is associated with an eigenvector
tangent to the $s$-axis), behaving as saddle. { For Bianchi I models the
eigenvalues for $R_3$ are the same, apart from the first one since the
  extra $\Omega_k$-direction    is not required for the analysis. Thus, the 
stability conditions of Bianchi I are the same as for LRS Bianchi III.}

For the curve of critical points $R_4$ the associated
eigenvalues are
\begin{eqnarray*}
&\left\{0,\frac{3 (\gamma -2)}{2},\frac{3 \gamma }{2}, -\frac{3}{4}
\left[2-\gamma +\sqrt{(2-\gamma ) (\gamma  (24 \gamma
   -41)+18)}\right], \right.\nonumber\\ & \left. -\frac{3}{4} \left[2-\gamma
-\sqrt{(2-\gamma ) (\gamma  (24 \gamma
   -41)+18)}\right]\right\},
\end{eqnarray*}
and thus they are
non-hyperbolic (the zero-eigenvalue is associated with an eigenvector
tangent to the $s$-axis), behaving as saddle. { This point does not belong
to the LRS Bianchi I boundary set.}

For the isolated critical point $R_5$  the  
eigenvalues write as
\begin{eqnarray*} & \left\{-2,-3,-3 \gamma ,\frac{1}{2} \left[-\sqrt{9-12
f(0)}-3\right], \frac{1}{2} \left[\sqrt{9-12
f(0)}-3\right]\right\},\end{eqnarray*} therefore it is stable for
$f(0)>0$ or   saddle for $f(0)<0$.  {For Bianchi I models the
eigenvalues for $R_5$ are the same, apart from the first one since the
  extra $\Omega_k$-direction    is not required for the analysis. Thus, the 
stability conditions of Bianchi I are the same as for LRS Bianchi III.}

For the critical point  $R_6(s^*)$ the  
eigenvalues read
\begin{eqnarray*}
&\left\{3 \gamma -2,-\frac{3}{2} (2-\gamma ),-\frac{3}{4} \left\{-\gamma
+\sqrt{(2-\gamma )
\left[\frac{24 \gamma
   ^2}{\left({s^*}\right)^2}-9 \gamma +2\right]}+2\right\},\right.\nonumber\\
& \left.
\frac{3}{4} \left\{-\gamma -\sqrt{(2-\gamma ) \left[\frac{24 \gamma
   ^2}{\left({s^*}\right)^2}-9 \gamma +2\right]}+2\right\}, -\frac{3 \gamma 
f'\left(s^*\right)}{s^*}\right\},
\end{eqnarray*}
that is it is asymptotically stable if
either $0<\gamma<\frac{2}{3},\, s^*<-\sqrt{3\gamma}$ and
$f'(s^*)<0,$ or $0<\gamma<\frac{2}{3},\, s^*>\sqrt{3\gamma}$
and $f'(s^*)>0$, otherwise it is a saddle point. {For the Bianchi I
subcase the eigenvalues associated to $R_6(s^*)$ are the same apart from the
first one. This leads to a broader interval of stability conditions than
that of Bianchi III, for which the new bifurcation value
$\gamma=\frac{2}{3}$ appears. Summarizing, for the LRS Bianchi I class,
$R_6(s^*)$ is stable for  
 $0<\gamma< 2,\, s^*<-\sqrt{3\gamma}$ and
$f'(s^*)<0,$   or  $0<\gamma< 2,\, s^*>\sqrt{3\gamma}$
and $f'(s^*)>0.$}

For the critical point $R_7(s^*)$ the  
eigenvalues are
\begin{eqnarray*}& \left\{-\frac{3}{4} \left\{2-\gamma +\frac{\sqrt{(2-\gamma
) \left[\Delta_2+4 \left(3 \gamma
   ^2+\sqrt{\Delta_1}\right)\right]}}{s^*}\right\}, \right.
\nonumber\\ & \left. 
	-\frac{3}{4} \left\{2-\gamma -\frac{\sqrt{(2-\gamma )
\left[\Delta_2+4 \left(3 \gamma
^2+\sqrt{\Delta_1}\right)\right]}}{s^*}\right\},\right.\nonumber\\ & \left.
-\frac{3}{4}
   \left\{2-\gamma +\frac{\sqrt{(2-\gamma ) \left[\Delta_2+4 \left(3 \gamma
^2-\sqrt{\Delta_1}\right)\right]}}{s^*}\right\}, \right.
\nonumber\\ & \left. 
-\frac{3}{4} \left\{2-\gamma -\frac{\sqrt{(2-\gamma ) \left[\Delta_2+4
\left(3 \gamma
^2-\sqrt{\Delta_1}\right)\right]}}{s^*}\right\},
-\frac{3 \gamma
   f'\left(s^*\right)}{s^*}\right\},
   \end{eqnarray*}
where
$\Delta_1={9 \gamma ^4+2 [3 \gamma  (3 \gamma -4)+2] \left({s^*}\right)^2
\gamma
^2+[\gamma  (3 \gamma -4)+2]^2 \left(s^*\right)^4}$ and $\Delta_2=[10-(25-12
\gamma )
\gamma] ({s^*})^2$.
Therefore, it is stable for either $\frac{2}{3}<\gamma
<1,s^*<-\sqrt{-\frac{\gamma }{\gamma -1}}, f'\left(s^*\right)<0$ or
$\frac{2}{3}<\gamma <1,
   s^*>\sqrt{-\frac{\gamma }{\gamma -1}}, f'\left(s^*\right)>0$, otherwise it
is a saddle point. { This point does not belong to the LRS Bianchi I
boundary set.}

For the critical point $R_8(s^*)$ the associated
eigenvalues are $$\left\{\left({s^*}\right)^2-2, \frac{1}{2}
\left[\left({s^*}\right)^2-6\right],\frac{1}{2}
\left[\left({s^*}\right)^2-6\right],\left({s^*}
\right)^2-3 \gamma ,-s^*
f'(s^*)\right\},$$ thus it is
 non-hyperbolic for
${s^*}^2\in \{0,2,6, 3\gamma\}$ or $f'(s^*)=0$. It is asymptotically stable
for $0<\gamma\leq \frac{2}{3}, -\sqrt{3\gamma}<s^*<0, f'(s^*)<0$ or
$0<\gamma\leq \frac{2}{3}, 0<s^*<\sqrt{3\gamma}, f'(s^*)>0$ or
$\frac{2}{3}<\gamma\leq 2, -\sqrt{2}<s^*<0, f'(s^*)<0$ or
$\frac{2}{3}<\gamma\leq 2, 0<s^*<\sqrt{2}, f'(s^*)>0$, otherwise it is a
saddle point. {For the Bianchi I
subcase the eigenvalues associated to $R_8(s^*)$  are the same apart from the
first one (corresponding to $\Omega_k$-direction). 
Thus, we have a broader
interval of stability conditions than that of Bianchi III, for which the
new bifurcation values $\gamma=\frac{2}{3}$ and $s^*=\pm \sqrt{2}$ appear.
Hence, the stability conditions for $R_8(s^*)$ in the case of LRS Bianchi I
are $0<\gamma\leq
2,\, -\sqrt{3\gamma}<s^*<0,f'(s^*)<0,$   or $0<\gamma\leq 2,\,
0<s^*<\sqrt{3\gamma},f'(s^*)>0$. In any other case it is a saddle point. }

For the critical points $R_9^\pm$ the  
eigenvalues are $\left\{4,0,3\mp
\sqrt{\frac{3}{2}} s^*,6-3 \gamma ,\mp \sqrt{6} f'(s^*)\right\}$.
Therefore, $R_9^+$ ( $R_9^-$) is unstable for $0\leq
\gamma<2$, $s^*<0$ ($s^*>0$), $f'\left(s^*\right)<0$
($f'\left(s^*\right)>0$), otherwise they are saddle points.
{For Bianchi I models the
eigenvalues associated to  $R_9^\pm$ are the same, apart from the first one,
and thus the 
stability conditions of Bianchi I are the same with those of LRS Bianchi
III.}

{For the critical points $R_{10}(s^*)$ the associated eigenvalues are
\begin{eqnarray*}&\left\{-\frac{3 
\left[ ({s^*})^2+2\right]}{2 \left[({s^*})^2+1\right]},-\frac{3  \left[(\gamma -1)
    ({s^*})^2+\gamma \right]}{ ({s^*})^2+1},-\frac{3  s^*
   f'\left(s^*\right)}{ ({s^*})^2+1},  \right.
\nonumber\\ & \left. 
	-\frac{3 \left[ ({s^*})^2+2+\sqrt{\left[2+(s^*)^2\right]\left[18-7 (s^*)^2\right]}\right]}{4
   \left[ ({s^*})^2+1\right]}, \right. \nonumber\\ &\left.
	-\frac{3 \left[ ({s^*})^2+2-\sqrt{\left[2+(s^*)^2\right]\left[18-7 (s^*)^2\right]}\right]}{4 \left[ ({s^*})^2+1\right]}\right\}\nonumber,
   \end{eqnarray*}
thus they are nonhyperbolic for $({s^*})^2\in \left\{0,2, \frac{\gamma}{1-\gamma}\right\}$ or $f'(s^*)=0.$  They are stable for 
$\frac{2}{3}<\gamma<1,
-\sqrt{\frac{\gamma}{1-\gamma}}<s^*<-\sqrt{2}, f'(s^*)<0$;  or $1\leq \gamma\leq 2,
s^*<-\sqrt{2}, f'(s^*)<0$; or    $\frac{2}{3}<\gamma<1, \sqrt{2}<s^*<\sqrt{\frac{\gamma}{1-\gamma}},
f'(s^*)>0$; or  $1\leq \gamma\leq 2, s^*>\sqrt{2},
f'(s^*)>0$. This point
does not belong to the LRS Bianchi I boundary set.}

Finally, for the curve of critical points $C(s^*)$  the associated
eigenvalues are
\begin{eqnarray*} &\left\{4-2 \cos (u), 0, 6-3 \gamma, 3-\sqrt{\frac{3}{2}}
\sin (u)
s^*, -\sqrt{6} \sin (u)
f'\left(s^*\right)\right\}.
\end{eqnarray*}
Thus, its points are normally hyperbolic and the stability is determined by
the sign of the non zero eigenvalues. Hence, they are unstable for
$0\leq
\gamma<2,0<u<\pi, s^*<\sqrt{6} \csc (u), f'\left(s^*\right)<0$ or  $0\leq
\gamma<2,\pi<u<2\pi, s^*>\sqrt{6} \csc (u), f'\left(s^*\right)>0$, otherwise
they are 
saddle points. {For Bianchi I models the
eigenvalues for $C(s^*)$ are the same, apart from the first one,
and thus the 
stability conditions of Bianchi I are the same with those of LRS Bianchi
III.}

\end{appendix}

%\section*{References}

\end{document}